\documentclass[11pt, a4paper]{article}
\pdfoutput=1 

\usepackage[height=8.85in,width=6.75in]{geometry}

\usepackage{graphicx,rotating}     %usepackage without driver option
\usepackage[bookmarksopen,colorlinks=true,linkcolor=light_blue,
citecolor=light_pink,urlcolor=dark_red,linktoc=all]{hyperref}

\usepackage{amsmath}
\usepackage{mathtools}
\usepackage{amsfonts}
\usepackage{amssymb}
\usepackage{slashed}
\usepackage{braket}
\usepackage{bm}
\usepackage{bbm}
\usepackage{cite}
\usepackage{comment}
\usepackage{xcolor}
\usepackage[utf8]{inputenc}
\usepackage[footnotesize]{caption}
\usepackage{bbold}

\usepackage{fourier}

\newcommand{\dd}{\mathrm{d}}

\newcommand{\abs}[1]{\left\vert {#1} \right\vert}
\newcommand{\FCB}{fermion current basis }
\newcommand{\FCBnsp}{fermion current basis}
\newcommand{\FCBsnsp}{fermion current bases}

\makeatletter
\@addtoreset{equation}{section}

\makeatother

\usepackage{multicol}
\definecolor{dark_red}{rgb}{0.7, 0., 0.}
\definecolor{light_pink}{rgb}{1,0.4,0.4}
\definecolor{light_blue}{rgb}{0.284602,0.317763,0.963947}
\definecolor{cred}{RGB}{180,50,40} 
\definecolor{darkgreen}{RGB}{0, 100, 0}
\definecolor{desy_blue}{HTML}{009EE2}
\definecolor{desy_orange}{HTML}{FD8800}
\definecolor{forestgreen}{HTML}{228B22}
\definecolor{ochre}{HTML}{CCAA2B}
\begin{document}

\hypersetup{pageanchor=false}
\begin{titlepage}

\begin{center}

\hfill CERN-TH-2022-069 \\
\hfill UMN-TH-4120/22 \\
\hfill FTPI-MINN-22-11 \\
\hfill KEK-TH-2415\\

\vskip 0.5in

{\Huge \bfseries 
Transient phenomena \\ \vspace{5mm} in the axion assisted Schwinger effect} \\
\vskip .8in

{\Large Valerie Domcke$^{a,b}$, Yohei Ema$^c$, Kyohei Mukaida$^{d,e}$}

\vskip .3in
\begin{tabular}{ll}
$^a$& \!\!\!\!\!\emph{Theoretical Physics Department, CERN, 1211 Geneva 23, Switzerland}\\
$^b$& \!\!\!\!\!\emph{Laboratory for Particle Physics and Cosmology, Institute of Physics, }\\[-.3em]
& \!\!\!\!\!\emph{School of Basic Sciences, EPFL, 1015 Lausanne, Switzerland}\\ 
$^c$& \!\!\!\!\!\emph{William I. Fine Theoretical Physics Institute, School of Physics and Astronomy, }\\[-.3em]
& \!\!\!\!\!\emph{ University of Minesota, Minneapolis, M 55455, USA} \\
$^d$& \!\!\!\!\!\emph{Theory Center, IPNS, KEK, 1-1 Oho, Tsukuba, Ibaraki 305-0801, Japan}\\
$^e$& \!\!\!\!\!\emph{Graduate University for Advanced Studies (Sokendai), }\\[-.3em]
& \!\!\!\!\!\emph{1-1 Oho, Tsukuba, Ibaraki 305-0801, Japan}
\end{tabular}

\end{center}
\vskip .6in

\begin{abstract}
\noindent
Particle production induced by a time-dependent background is well understood as the projection of the time-evolved initial state onto a set of final states. While the asymptotic initial and final states are well defined in the usual way, the definition of particles and antiparticles at intermediate times in the presence of external fields is ambiguous. These external fields moreover induce divergences which require regularization.
In this paper we clarify some subtleties in the computation of transient effects in physical quantities for fermions in a homogeneous axion background, including Schwinger production in background electromagnetic fields. The presence of the axion requires particular care as well as knowledge of the UV theory when regulating the theory and computing the vacuum contribution to the fermion energy.
\end{abstract}

\end{titlepage}

\tableofcontents
\thispagestyle{empty}
\renewcommand{\thepage}{\arabic{page}}
\renewcommand{\thefootnote}{$\natural$\arabic{footnote}}
\setcounter{footnote}{0}
\newpage
\hypersetup{pageanchor=true}

%%%%%%%%%%%%%
\section{Introduction}
\label{sec:intro}
%%%%%%%%%%%%%

The axion was first introduced as the angular degree of freedom of an approximate global $U(1)$ symmetry to address the strong CP problem in QCD~\cite{Peccei:1977hh,Peccei:1977ur,Weinberg:1977ma,Wilczek:1977pj}. In modern cosmology, this concept has been generalized to axion-like particles which have been postulated to play a key role in string theory~\cite{Banks:2003sx,Svrcek:2006yi,Ibanez:1986xy}, inflation~\cite{Freese:1990rb,Anber:2006xt}, baryogenesis~\cite{Cohen:1987vi,Cohen:1988kt}, dark matter~\cite{Preskill:1982cy,Abbott:1982af,Dine:1982ah} and in cosmological approaches to the hierarchy problem~\cite{Graham:2015cka}. At the level of quantum field theory, these axion-like particles (axions for short in following) are pseudoscalars which couple to fermions and gauge fields only through derivative couplings. Of particular phenomenological interest are couplings to the fermions and gauge groups of the Standard Model (SM), in particular the electromagnetic sector.\footnote{
At energies above the electroweak phase transition, the relevant Abelian gauge group is hypercharge. For simplicity, we will use the notation familiar from electromagnetism throughout this paper, though of course the discussion applies to any Abelian gauge group with charged fermions. For non-Abelian gauge groups, see \emph{e.g.}~\cite{Domcke:2018gfr, Adshead:2022ecl}.
}
In this case, the gauge fields, charged fermions and the axion form a coupled system linked by non-linear interactions. For example, a non-vanishing axion velocity can trigger a tachyonic instability in one of the gauge field helicity modes~\cite{Turner:1987bw,Garretson:1992vt}, Schwinger pair production of charged fermions in background gauge fields~\cite{Heisenberg:1935qt,Schwinger:1951nm} leads to an induced current which dampens this tachyonic instability~\cite{Domcke:2018eki,Domcke:2019qmm,Fujita:2022fwc}, and the gauge fields in turn back-react as an effective friction to the axion motion~\cite{Anber:2009ua}. Moreover, it was recently pointed out that a non-vanishing axion velocity has the intriguing property of boosting the Schwinger pair production of charged fermions in an electric field~\cite{Domcke:2021fee}. In the \textit{axion-assisted Schwinger effect}~\cite{Domcke:2021fee} a sufficiently large constant axion velocity can exponentially enhance the Schwinger production rate for finite fermion momentum in the final state whereas the \textit{axion-driven pair production}~\cite{Kitamoto:2021wzl} relies a non-vanishing acceleration of the axion field.

In the literature, particle production is usually discussed in terms of the asymptotic initial and final states, where the particles and antiparticles are well defined.
Starting in a ground state in the asymptotic past, the electromagnetic and/or axion background fields are switched on for some finite time, and the final state in the asymptotic future, after switching off the background fields, is determined by solving the equations of motion for the Bogoliubov coefficients.
The time-dependence of the background fields, \emph{i.e.} the mismatch of the ground state in the asymptotic past and future, is the origin of the fermion production.
On the other hand, transient effects occurring when the external fields are active are more subtle.
At this intermediate stage, the definition of particles and antiparticles is ambiguous and
the divergences associated with the external fields call for an appropriate regularization scheme.
In a practical computation, one may select a particular basis of creation/annihilation operators, which however results in 
seemingly very different transient effects for different choices of basis, as is well known in the case of Schwinger production in time-dependent electric fields~\cite{Dabrowski:2014ica,Dabrowski:2016tsx}.
However, physical quantities such as the fermion current and energy density should not depend on such unphysical reparametrizations of the theory.

In this paper, we extend the existing results in a three-fold way. Firstly, we point out that in order to compute backreaction effects in the coupled axion, gauge field and fermion system, it is insufficient to focus on results for the asymptotic future, it becomes instead crucial to correctly capture transient phenomena. 
We demonstrate explicitly that including vacuum contributions and an appropriate regularization scheme, the apparent differences caused by the choice of basis are merely differences in interpretation, \emph{i.e.}\ labeling terms as vacuum contribution, time-dependent Bogoliubov coefficients or higher-dimensional operators, 
while the physical observables are unaffected by these different basis choices -- as they should. We secondly use this opportunity to compare different approaches in the existing literature, in particular~\cite{Domcke:2021fee},\cite{Kitamoto:2021wzl} and \cite{Adshead:2021ezw}. In this context, we highlight the importance of choosing a regularization scheme which respects the symmetries of the theory and the benefits of making use of the Adler-Bell-Jackiw anomaly equation~\cite{Adler:1969gk,Bell:1969ts} as a necessary verification of this. Thirdly, we numerically and analytically compute the transient contributions to fermion production in the presence of a background axion field.
We show that \emph{e.g.}\ the fermion energy in the limit of large axion velocities can be exponentially enhanced, even in the absence of electromagnetic background fields. However, we note that the dominant contribution can be interpreted as dimension-eight operator in the axion effective theory.
This proves that the transient effects in the axion-fermion system cannot be reliably computed without knowledge of the UV properties of the theory, \textit{i.e.}\ a particular realization of the Peccei Quinn theory.

The remainder of this paper is organized as follows. After reviewing the key ingredients of the axion assisted Schwinger effect in Section~\ref{sec:review}, we discuss in some detail different basis choices and regularization schemes in Section~\ref{sec:basis}. The computation of transient phenomena in observable quantities is presented in Section~\ref{sec:transients}, including a discussion on the UV sensitivity. We conclude in Sec.~\ref{sec:conclusions}. Various technical details are relegated to the appendices. App.~\ref{app:notation} specifies our notation and conventions, while App.~\ref{app:details} gives the necessary details on the solutions of the Dirac equation.

%%%%%%%%%%%%%
\section{Fermion production in a gauge field and axion background}
\label{sec:review}
%%%%%%%%%%%%%

In this section we review the axion assisted Schwinger effect~\cite{Domcke:2021fee}.
We consider the following action:
\begin{align}
	S = \int \dd^4x \left[\frac{1}{2}\left(\partial \phi\right)^2 - V(\phi) - \frac{1}{4}F_{\mu\nu}F^{\mu\nu}
	+ \bar{\psi}\left(i\slashed{D} - m e^{2i c_m \phi/f_a \gamma_5}\right)\psi
	+ c_A \frac{\alpha}{4\pi f_a} \phi F_{\mu\nu}\tilde{F}^{\mu\nu}
	+ c_5 \frac{\partial_\mu \phi}{f_a}\bar{\psi}\gamma^\mu \gamma_5 \psi
	\right],
	\label{eq:full_action}
\end{align}
where $\phi$ is the axion field with its potential $V(\phi)$ and its decay constant $f_a$, 
$A_\mu$ is the U(1) gauge field with its field strength $F_{\mu\nu}$, 
and $\psi$ is a fermion with the mass $m$ whose production we will discuss in detail. 
The covariant derivative is defined by $D_\mu = \partial_\mu + ig Q A_\mu$,
where $g$ is the gauge coupling with $\alpha = g^2/4\pi^2$ and $Q$ is the charge of the fermion.
The dual field strength tensor is defined as $\tilde{F}_{\mu\nu} = \epsilon_{\mu\nu\rho\sigma}F^{\rho\sigma}/2$ with $\epsilon^{0123} = +1$.
Finally, $c_m$, $c_A$ and $c_5$ are the coupling constants between the axion and the gauge fields/fermions.
The Dirac equation reads
\begin{align}
	\left[i \slashed{D} - m e^{2i\theta_m \gamma_5} + \partial_\mu \theta_5
	\gamma^\mu \gamma_5\right] \psi = 0,
	\label{eq:dirac_eq}
\end{align}
where we denote $\theta_i = c_i \phi/f_a$.
We solve this equation in the presence of background electromagnetic fields and 
a homogeneous axion field with $\dot{\phi} \neq 0$.

It is well known that a strong electric field creates pairs of fermions and anti-fermions through the Schwinger effect~\cite{Heisenberg:1935qt,Schwinger:1951nm}.
The production rate $\Gamma$ is proportional to
\begin{align}
	\Gamma \propto \exp\left[-\frac{\pi m_T^2}{g\abs{Q}E}\right],
	\quad
	m_T^2 = m^2 + p_T^2,
\end{align}
where $p_T$ is the transverse momentum, \emph{e.g.} $p_T^2 = p_x^2 + p_y^2$ 
if the electric field is in $\hat{z}$-direction.
Note that beyond the fermion mass suppression,
the production of the modes with large transverse momentum is exponentially further suppressed.
As however recently demonstrated in~\cite{Domcke:2021fee} 
this suppression from the transverse momentum disappears once 
the axion velocity is large enough, and the rate eventually becomes
\begin{align}
	\Gamma \propto \exp\left[-\frac{\pi m^2}{g\abs{Q}E}\right]
	~~~~~\mathrm{for}~~~~~
	\dot{\theta}_{5+m}^2 \gtrsim \frac{\pi m^2 p_T^2}{g\abs{Q}E},
	\label{eq:beta2_axion_assisted}
\end{align}
where $\theta_{5+m} = (c_5+c_m)\phi/f_a$.
This enhances the Schwinger effect for non-zero transverse momentum 
and was hence dubbed the axion assisted Schwinger effect.
The purpose of this section is to summarize the key ingredients of this effect 
(see~\cite{Domcke:2021fee} and App.~\ref{app:details} for more details).

In anticipation of our later discussion, we note that Eq.~\eqref{eq:dirac_eq} has a redundancy.
Indeed,
the couplings shift under the chiral rotation $\psi \to e^{ic\phi/f_a \gamma_5}\psi$ as
\begin{align}
	c_5 \to c_5 - c,
	\quad
	c_m \to c_m + c.
\end{align}
This indicates that only the combination $c_5 + c_m$ is physical,
and one can eliminate either $c_5$ or $c_m$ by a chiral rotation.
In this section we focus on the $c_5 = 0$ basis following~\cite{Domcke:2021fee},
and comment on the $c_m = 0$ basis at the end of Sec.~\ref{subsec:axion_schwinger}.
A particular basis choice implies a natural definition of particles and antiparticles by the corresponding positive and negative frequency modes.
Hence, throughout this paper, we identify the basis choice with respect to the redundancy under a chiral rotation with a particular choice of creation/annihilation operators, although these two concepts are not equal, strictly speaking.
This basis transformation is a key topic of Sec.~\ref{sec:basis}.

%%%%%%%%%%%%%
\subsection{Dirac equation with electric field and axion}
\label{subsec:dirac_eq_E}
%%%%%%%%%%%%%

In this section, we focus on the case with a background electric field and axion velocity, 
without any magnetic field.
We consider background gauge and axion fields parameterized by
\begin{align}
	A^{\mu} = \left(0, 0, 0, A_z(t)\right),
	\qquad
	\phi = \phi(t),
\end{align}
where the electric field is given by $E = -\dot{A}_z $.
Without loss of generality we take $E \geq 0$.
We will take this background to be constant at some initial and final time, $\dot \phi = 0$, $\dot A = 0$, with a time-dependence at intermediate times.
The Dirac equation is then given by
\begin{align}
	0 &= \begin{pmatrix} i\partial_0 + \Pi_z - \dot{\theta}_5 & p_x - ip_y & -me^{2i\theta_m} & 0 \\
	p_x + ip_y & i\partial_0 - \Pi_z - \dot{\theta}_5 & 0 & -me^{2i\theta_m} \\
	-me^{-2i\theta_m} & 0 & i\partial_0 - \Pi_z + \dot{\theta}_5 & -(p_x - ip_y) \\
	0 & -me^{-2i\theta_m} & -(p_x + ip_y) & i\partial_0 + \Pi_z + \dot{\theta}_5
	\end{pmatrix}
	\psi.
\end{align}
We solve this equation with the Bogoliubov coefficient method,
which describes the mixing of the positive and negative frequency modes.
The Bogoliubov coefficients $\alpha_\lambda$ and $\beta_\lambda$ 
evolve as (see App.~\ref{app:details} for derivation)
\begin{align}
	\begin{pmatrix}
	\dot{\alpha}_1 \\
	\dot{\alpha}_2 \\
	\dot{\beta}_1 \\
	\dot{\beta}_2 
	\end{pmatrix}
	=&
	\left[
	i\dot{\theta}_{5+m}
	\begin{pmatrix}
	- \frac{m}{m_T}\frac{\Pi_z}{\Omega} & \frac{p_T}{m_T} & \frac{m}{\Omega}e^{2i\Theta} & 0 \\
	\frac{p_T}{m_T} & \frac{m}{m_T}\frac{\Pi_z}{\Omega} & 0 & -\frac{m}{\Omega}e^{2i\Theta} \\
	\frac{m}{\Omega}e^{-2i\Theta} & 0 & \frac{m}{m_T}\frac{\Pi_z}{\Omega} & \frac{p_T}{m_T} \\
	0 & -\frac{m}{\Omega}e^{-2i\Theta} & \frac{p_T}{m_T} & -\frac{m}{m_T}\frac{\Pi_z}{\Omega}
	\end{pmatrix}
	%\right. \nonumber \\ &~~~~~~~~~~~ \left.
	+ \frac{m_T \dot{\Pi}_z}{2\Omega^2}
	\begin{pmatrix}
	0 & 0 & -e^{2i\Theta} & 0 \\
	0 & 0 & 0 & -e^{2i\Theta} \\
	e^{-2i\Theta} & 0 & 0 & 0 \\
	0 & e^{-2i\Theta} & 0 & 0
	\end{pmatrix}
	\right]
	\begin{pmatrix}
	{\alpha}_1 \\
	{\alpha}_2 \\
	{\beta}_1 \\
	{\beta}_2 
	\end{pmatrix},
	\label{eq:alphabeta_woB}
\end{align}
where $\Pi_z = p_z - gQ A_z$, $p_T^2 = p_x^2 + p_y^2$, $m_T^2 = p_T^2 + m^2$,
$\Omega = \sqrt{\Pi_z^2 + m_T^2}$, and $\Theta = \int^t \dd t \,  \Omega$.
We solve these equations with two different initial conditions,
\begin{align}
	\alpha_{1}^{(1)} &= 1,
	\quad
	\alpha_{2}^{(1)} = \beta_{1}^{(1)} = \beta_{2}^{(1)} = 0,
	\qquad\mathrm{and}\qquad
	\alpha_{2}^{(2)} = 1,
	\quad
	\alpha_{1}^{(2)} = \beta_{1}^{(2)} = \beta_{2}^{(2)} = 0.
\end{align}
Using the Bogoliubov coefficients, we quantize the fermion as
\begin{align}
	\psi &= \int\frac{\dd^3 p}{(2\pi)^3}e^{i\vec{p}\cdot\vec{x}}
	e^{i\theta_5 \gamma_5}\sum_{\lambda=1,2}\left[
	B_\lambda u_\lambda e^{-i\Theta}
	+ D_\lambda^\dagger v_\lambda e^{+i\Theta}\right],
\end{align}
where $u_\lambda$ and $v_\lambda$ are the solutions of the Dirac equation
for the constant background fields.
Note that we have extracted the factor $e^{i\theta_5 \gamma_5}$ to eliminate the derivative coupling between the axion and the fermion current, \emph{i.e.}\ $c_5 = 0$ in this basis.
The creation and annihilation operators at the intermediate times, $B_\lambda$ and $D_\lambda$,
are expressed as
\begin{align}
	B_\lambda &= \sum_{\lambda'=1,2}
	\left[\alpha_\lambda^{(\lambda')}b_{\lambda'} 
	- (-)^{\lambda+\lambda'}{\beta_{\lambda}^{(\lambda')}}^{*} d_{\lambda'}^\dagger\right],
	\quad
	D_\lambda^\dagger = \sum_{\lambda'=1,2}
	\left[\beta_\lambda^{(\lambda')}b_{\lambda'} + (-)^{\lambda+\lambda'}
	{\alpha_\lambda^{(\lambda')}}^* d_{\lambda'}^\dagger\right],
\end{align}
where $b_\lambda$ and $d_\lambda$ are the creation and annihilation operators in the infinite past.
We impose the anti-commutation relation as
\begin{align}
	\left\{b_{\lambda_1}(\vec{p}), b_{\lambda_2}^\dagger(\vec{p}')\right\}
	&= \left\{d_{\lambda_1}(\vec{p}), d_{\lambda_2}^\dagger(\vec{p}')\right\}
	=(2\pi)^3\delta_{\lambda_1 \lambda_2}\delta^{(3)}(\vec{p}-\vec{p}'),
	\\
	\left\{b_{\lambda_1}(\vec{p}), d_{\lambda_2}^\dagger(\vec{p}')\right\}
	&= \left\{b_{\lambda_1}(\vec{p}), d_{\lambda_2}(\vec{p}')\right\}
	= 0.
\end{align}
It then follows from Eq.~\eqref{eq:alphabeta_woB} that $B_\lambda$ and $D_\lambda$
satisfy the same equal time anti-commutation relations, \emph{i.e.},
\begin{align}
	\left\{B_{\lambda_1}(t, \vec{p}), B_{\lambda_2}^\dagger(t, \vec{p}')\right\}
	&= \left\{D_{\lambda_1}(t, \vec{p}), D_{\lambda_2}^\dagger(t, \vec{p}')\right\}
	=(2\pi)^3\delta_{\lambda_1 \lambda_2}\delta^{(3)}(\vec{p}-\vec{p}'),
	\\
	\left\{B_{\lambda_1}(t, \vec{p}), D_{\lambda_2}^\dagger(t, \vec{p}')\right\}
	&= \left\{B_{\lambda_1}(t, \vec{p}), D_{\lambda_2}(t, \vec{p}')\right\}
	= 0.
\end{align}
%%

%%%%%%%%%%%%%
\subsection{Axion assisted Schwinger effect}
\label{subsec:axion_schwinger}
%%%%%%%%%%%%%

%%
\begin{figure}[t]
	\centering
 	\includegraphics[width=0.495\linewidth]{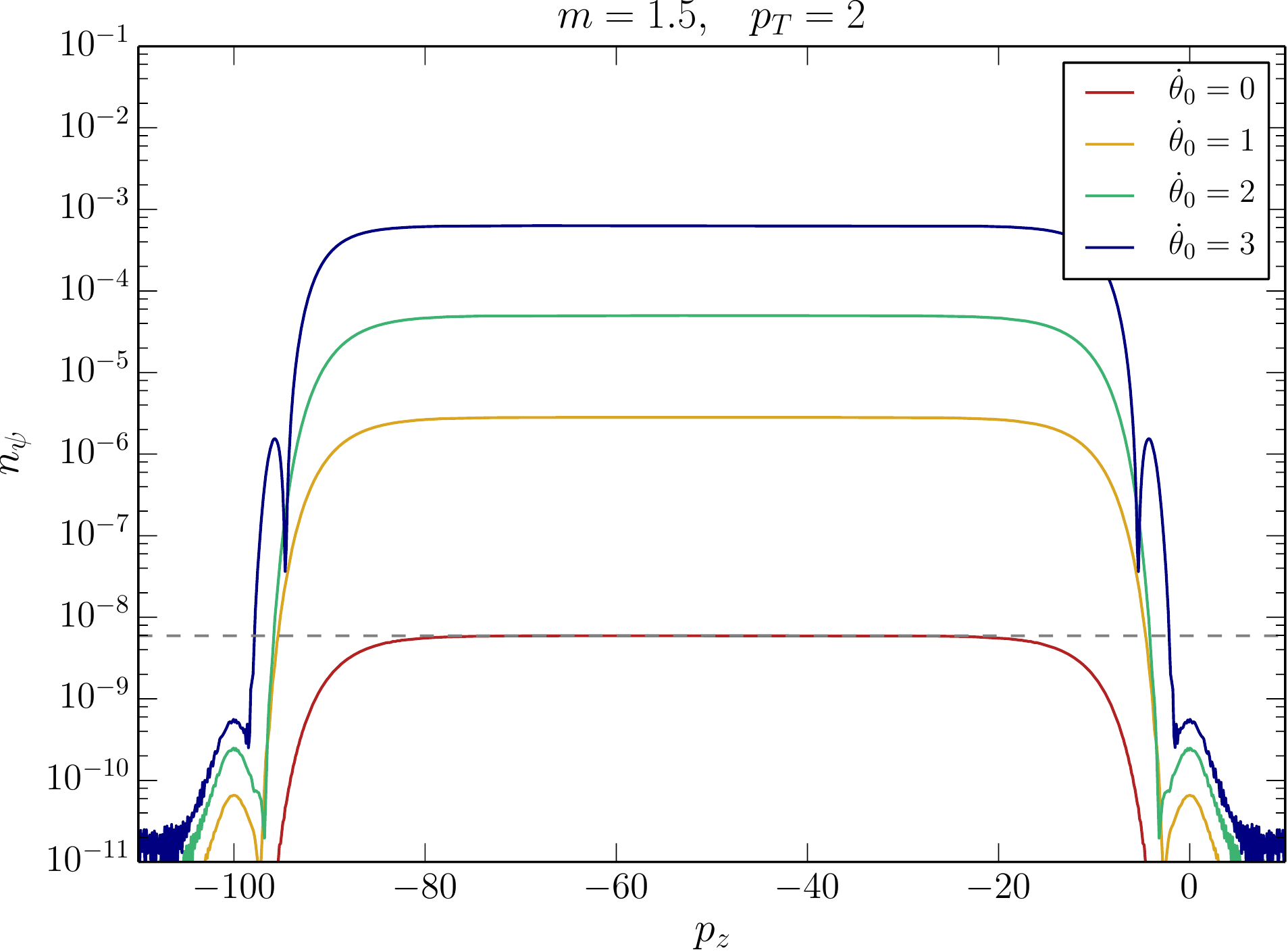}
 	\includegraphics[width=0.495\linewidth]{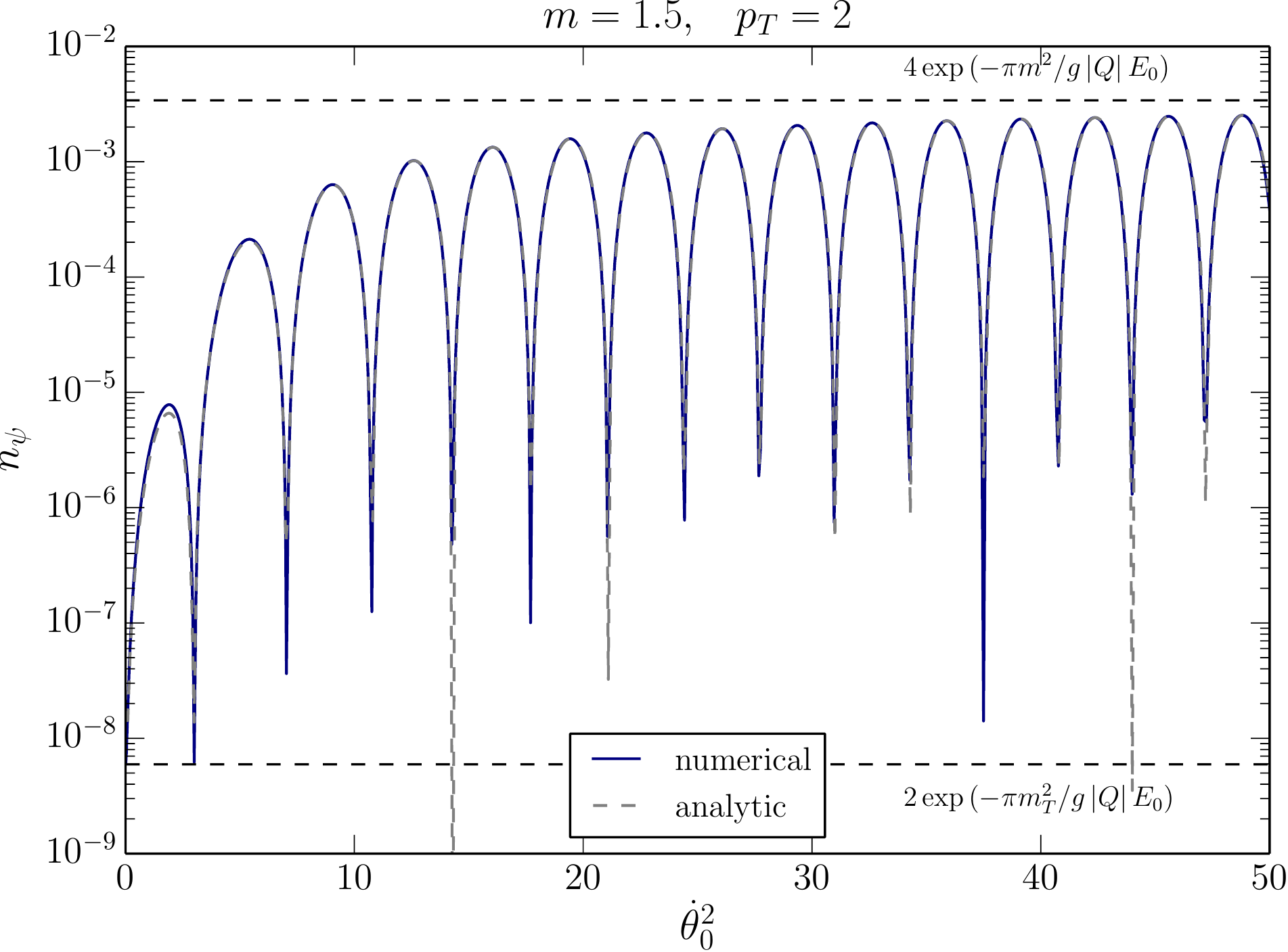}
	\caption{\small Numerical solutions for the total number of produced particles after turning off the background fields.
	\emph{Left:} spectrum of produced particles for different values of the background axion velocity $\dot{\theta}_0$ as a function of the fermion momentum $p_z$.
	The gray dashed line corresponds to the standard Schwinger effect formula,
	$n_\psi = 2 \exp(-\pi m_T^2/g\abs{Q}E_0)$, where the factor 2 accounts for the chirality.
	\emph{Right:} the height of the spectrum (at $p_z = -50 \sqrt{g\abs{Q}E_0}$) 
	as a function of $\dot{\theta}_0^2$.
	The blue line corresponds to the numerical solution of the time evolution equation of the Bogoliubov coefficients,
	while the gray dashed line shows the analytical expression~\eqref{eq:spectrum_analytic}. 
	In both figures, we take $m = 1.5$ and $p_T = 2$ in units of $g\abs{Q}E_0 = 1$.}
	\label{fig:axion_schwinger}
\end{figure}

We now turn to the fermion pair production in the presence of a background electric field and axion velocity.
We focus on the axion assisted Schwinger effect, and thus take the axion velocity to be constant.\footnote{
	If the axion velocity is exactly constant, the axion field by itself cannot produce fermions, it can only enhance the production
	driven by the electric field. In other words, the production rate vanishes for $E = 0$ for constant $\dot{\phi}$.
	If $\ddot{\phi} \neq 0$, the axion can produce fermions independently of the background gauge field,
	which includes the standard particle production by an oscillating axion background field decaying into fermions as well as axion-driven pair production. 
}
It is well known that the notion of particles is ill-defined in the time-dependent background,
and thus we turn on and off the external electric and axion fields adiabatically at $t_\text{min}$ and $t_\text{max}$, respectively. 
To be specific, we take the background fields as
\begin{align}
	\dot{\theta}_{5+m} &= \frac{\dot{\theta}_0}{4}\left[1+\mathrm{tanh}\left(\frac{t-t_\mathrm{min}}{T}\right)\right]
	\left[1-\mathrm{tanh}\left(\frac{t-t_\mathrm{max}}{T}\right)\right],
	~~
	\dot{A}_z = -\frac{E_0}{4}\left[1+\mathrm{tanh}\left(\frac{t-t_\mathrm{min}}{T}\right)\right]
	\left[1-\mathrm{tanh}\left(\frac{t-t_\mathrm{max}}{T}\right)\right],
\end{align}
and start to solve the system well before $t_\mathrm{min}$.
Here we restrict ourselves to evaluate 
the particle number $n_\psi = \sum_{\lambda, \lambda'}\vert \beta_\lambda^{(\lambda')}\vert^2$ 
only well after $t_\mathrm{max}$.
We will study physical observables at the intermediate times in the subsequent sections.

In Fig.~\ref{fig:axion_schwinger} we show a numerical solution of the equation of motion.
We take $t_\mathrm{min} = 0$, $t_\mathrm{max} = 100\sqrt{g\abs{Q}E_0}$, $T=\sqrt{50/g\abs{Q}E_0}$, and evaluated
the spectrum at $t = 1.5 t_\mathrm{max}$.
In the left panel, we show the spectrum for $Q=+1$, $m = 1.5$ and $p_T = 2$ in the unit $g\abs{Q}E_0 = 1$ 
for different values of $\dot{\theta}_0$
(see~\cite{Domcke:2021fee} for several other values of $p_T$ and $m$).
The enhanced part of the spectrum corresponds to the modes that cross $\Pi_z = 0$.
Since the gap size between the positive and negative frequency modes 
is smallest at $\Pi_z = 0$ (for $\dot{\theta}_0 < p_T$), the modes are most easily excited at this point.
The height of the plateau is enhanced for non-zero $\dot{\theta}_0$ compared to the standard Schwinger effect formula,
$n_\psi = 2 \exp(-\pi m_T^2/g\abs{Q}E_0)$ (which we show as the dashed line), 
corresponding to the axion assisted Schwinger effect.
In order to take a closer look at this enhancement, in the right panel,
we plot the height of the plateau versus $\dot{\theta}_0^2$.
As one can see, the envelope eventually approaches to $\exp(-\pi m^2/ g\abs{Q}E_0)$
for large enough $\dot{\theta}_0$.
In other words, the suppression from the transverse momentum is absent for a sufficiently large axion velocity.
On top of this enhancement, the height of the plateau strongly oscillates as a function of $\dot{\theta}_0$.
As demonstrated in~\cite{Domcke:2021fee} this result is well reproduced by the analytical expression\footnote{
	Here we assume that $\dot{\theta}_0$ is positive.
	If $\dot{\theta}_0$ is negative, we should replace $\Omega_-$ by $\Omega_+$.
}
\begin{align}
	n_\psi = \abs{\exp\left[2i\int_0^{\Pi_{++}} \frac{\dd\Pi_z}{g\abs{Q}E_0} \Omega_-\right]}^2
	+ \abs{\exp\left[2i\int_0^{\Pi_{--}}\frac{\dd\Pi_z}{g\abs{Q}E_0} \Omega_-\right]}^2
	- 2\mathrm{Re}\left[\exp\left[
	2i\int_{\Pi_{-+}}^{\Pi_{++}}\frac{\dd\Pi_z}{g\abs{Q}E_0} \Omega_-
	\right]\right],
	\label{eq:spectrum_analytic}
\end{align}
which we show as the gray line in the right panel.
Here we have defined
\begin{align}
	\Omega_\pm &= \sqrt{\left(\sqrt{\Pi_z^2 + p_T^2}\pm\dot{\theta}_{0}\right)^2 + m^2},
	\quad
	\Pi_{\lambda \lambda'}
	= \lambda\sqrt{(\dot{\theta}_0 + i\lambda' m)^2 - p_T^2}.
\end{align}
The latter corresponds to the solution of $\Omega_-(\Pi_z) = 0$ in the complex $\Pi_z$-plane.
This suggests that the dispersion relation $\Omega_-$ is crucial for the axion assisted Schwinger effect.
Indeed, as shown in~\cite{Domcke:2021fee}, 
this dispersion relation arises due to spin-momentum interaction
induced by the axion in the non-relativistic limit of the fermion.

Ref.~\cite{Kitamoto:2021wzl} confirms this result in the $c_m = 0$ basis.
Indeed, one can eliminate $c_m$ by the chiral rotation $\psi \to e^{-i\theta_m \gamma_5}\psi$.
The Bogoliubov coefficients in this basis satisfy
\begin{align}
	\begin{pmatrix}
	\dot{\tilde\alpha}_+ \\
	\dot{\tilde\alpha}_- \\
	\dot{\tilde\beta}_+ \\
	\dot{\tilde\beta}_-
	\end{pmatrix}
	&= \begin{pmatrix}
	0 & Ae^{i(\Theta_+ - \Theta_-)} & -B_+ e^{2i\Theta_+} & -C e^{i(\Theta_+ + \Theta_-)} \\
	-A e^{-i(\Theta_+ - \Theta_-)} & 0 & -Ce^{i(\Theta_+ + \Theta_-)} & -B_-e^{2i\Theta_-} \\
	B_+ e^{-2i\Theta_+} & C e^{-i(\Theta_+ + \Theta_-)}& 0 & A e^{-i(\Theta_+ - \Theta_-)}  \\
	C e^{-i(\Theta_+ + \Theta_-)} & B_- e^{-2i\Theta_-} & -A e^{i(\Theta_+ - \Theta_-)}  & 0
	\end{pmatrix}
	\begin{pmatrix}
	\tilde{\alpha}_+ \\
	\tilde{\alpha}_- \\
	\tilde{\beta}_+ \\
	\tilde{\beta}_-
	\end{pmatrix},
\end{align}
where
\begin{align}
	\Theta_\pm = \int^t \Omega_\pm \dd t.	
\end{align}
See App.~\ref{app:details} for the explicit forms of $A, B$ and $C$, which are irrelevant for our purpose here.
The key point is that the above equation explicitly involves $\Omega_\pm$,
and thus it is more straightforward to see the importance of $\Omega_\pm$
on particle production with, \emph{e.g.}\ the WKB method~\cite{Kitamoto:2021wzl}.
However, this basis has its own subtlety, related to the regularization and renormalization,
and demonstrating this point is our main goal in Sec.~\ref{sec:basis}.
There we will see that each basis has its own advantages and pitfalls; the WKB analysis of particle production
is more straightforward in the $c_m = 0$ basis, while the $c_5=0$ basis is more suitable for 
regularizing and renormalizing the theory.

%%%%%%%%%%%%%
\section{Basis choice and Regularization}
\label{sec:basis}
%%%%%%%%%%%%%
As shown above, the axion assisted Schwinger effect may be more transparent
in the basis $c_m=0$ (which we will refer to as \emph{the \FCBnsp}) than in the basis $c_5=0$ 
(which we will call \emph{the Hamiltonian basis}).
However, there is a subtlety in the \FCB related to regularization.
Physical observables are independent of the choice of basis 
and in principle one can of course work in any basis.
In reality, however, physical observables are often divergent and require regularization and renormalization.
A regularization scheme should preserve the symmetries of the theory as much as possible, 
which is straightforward in the Hamiltonian basis  while can be more subtle in the \FCBnsp. 
The key point is that the Hamiltonian basis diagonalizes the Hamiltonian (as the name suggests),
while the \FCB does not.
The purpose of this section is to demonstrate this point in detail, by taking the anomaly equation as an example.

We emphasize that the regularization and renormalization are not only of academic interest.
If we focus only on physical quantities in the asymptotic future, after turning off the external fields 
(as we did in Sec.~\ref{sec:review}), 
there are no complications.
However, if we follow the time evolution of the physical quantities at intermediate times,
we necessarily have to deal with divergences, regularization and renormalization. 
Since physical quantities such as the induced current and the axial charge
at intermediate times are crucial to study, \emph{e.g.}\ the backreaction of the fermion production
to the axion and gauge field dynamics, a proper treatment of the renormalization and regularization is
phenomenologically relevant.

%%%%%%%%%%%%%
\subsection{Basis choice and Hamiltonian}
%%%%%%%%%%%%%
A fermion chiral rotation $\psi \to e^{ic\gamma_5}\psi$ shifts the couplings as $c_5 \to c_5 - c$ and $c_m \to c_m + c$.
This redundancy corresponds to the basis choice of the fermion that one works with.
In Ref.~\cite{Domcke:2021fee} we take the Hamiltonian basis ($c_5 =0$),
while Ref.~\cite{Kitamoto:2021wzl} takes the \FCB ($c_m = 0$).\footnote{
	More generally, we can take a basis with both non-zero $c_5$ and $c_m$. 
	This choice simply makes the equations more involved and we do not see any benefit, 
	and thus we focus only on the Hamiltonian and \FCB in this paper.
}
Our goal here is to see that
the Hamiltonian is diagonalized only in the Hamiltonian basis, and not in the \FCBnsp,
which is crucial for our study of the anomaly equation in Sec.~\ref{subsec:anomaly}.
Since we will discuss the anomaly equation,
we now include the magnetic field (anti-)parallel to the electric field.\footnote{
	Gauge field production towards the end of axion inflation provides this configuration~\cite{Anber:2006xt},
	so the inclusion of the magnetic field also has a phenomenological motivation.
} We take the background fields as
\begin{align}
	A^{\mu} = \left(0, 0, B x, A_z\right),
	\qquad
	\phi = \phi(t),
\label{eq:backgroundB}	
\end{align}
which corresponds to $\vec{E} = E\hat{z} = -\dot{A}_z \hat{z}$ and $\vec{B} = B \hat{z}$,
where the magnetic field $B$ is taken constant.
This configuration has
\begin{align}
	F_{\mu\nu}\tilde{F}^{\mu\nu} = -4\vec{E}\cdot \vec{B} \neq 0,
\end{align}
which makes the anomaly equation non-trivial at the intermediate times.

In the presence of the magnetic field,
a charged particle follows a circular trajectory in the plane orthogonal to the magnetic field.
This transverse motion is quantized in quantum mechanics, resulting in discrete Landau levels.
The mode without transverse motion is called the lowest Landau level,
while those with transverse motion are called higher Landau levels.
Therefore, in the presence of a magnetic field, the fermion is quantized as
\begin{align}
	\psi =& \int \frac{\dd p_y \dd p_z}{(2\pi)^2}e^{i(p_y y + p_z z)}
	e^{i\gamma_5 \theta_5}
	\left[B_0 u_0 e^{-i\Theta_0}
	+ D_0^\dagger v_0 e^{i\Theta_0}
	+ \sum_{n, \lambda=1,2}
	\left[ B_{n,\lambda} u_{n,\lambda} e^{-i\Theta_n}
	+
	D_{n,\lambda}^\dagger v_{n,\lambda} e^{i\Theta_n}
	\right]
	\right],
	\label{eq:quantization_Hamiltonian}
\end{align}
where
\begin{align}
	B_0 &= \alpha_0 b_0 - \beta_0^* d_0^\dagger,
	\quad
	B_{n,\lambda} = \sum_{\lambda'=1,2}\left[\alpha_{n,\lambda}^{(\lambda')} b_{n,\lambda'} - 
	\left(-1\right)^{\lambda+\lambda'}{\beta_{n,\lambda}^{(\lambda')}}^* d_{n,\lambda'}^\dagger\right],
	\\
	D_0^\dagger &= \beta_0 b_0 + \alpha_0^* d_0^\dagger,
	\quad
	D_{n,\lambda}^\dagger =
	\sum_{\lambda'=1,2} \left[\beta_{n,\lambda}^{(\lambda')} b_{n,\lambda'} 
	+ \left(-1\right)^{\lambda+\lambda'}{\alpha_{n,\lambda}^{(\lambda')}}^* d_{n,\lambda'}^\dagger \right],
\end{align}
in the Hamiltonian basis, and
\begin{align}
	\psi =& \int \frac{\dd p_y \dd p_z}{(2\pi)^2}e^{i(p_y y + p_z z)}
	e^{-i\gamma_5 \theta_m}
	\left[\tilde{B}_0 \tilde{u}_0 e^{-i\Theta_{0,s}}
	+ \tilde{D}_0^\dagger \tilde{v}_0 e^{i\Theta_{0,s}}
	+ \sum_{n, \lambda=\pm}\left[\tilde{B}_{n,\lambda}\tilde{u}_{n,\lambda} e^{-i\Theta_{n,\lambda}}
	+ \tilde{D}_{n,\lambda}^\dagger \tilde{v}_{n,\lambda} e^{i\Theta_{n,\lambda}}
	\right]
	\right],
	\label{eq:quantization_derivative}
\end{align}
where
\begin{align}
	\tilde{B}_0 &= \tilde{\alpha}_0 {b}_0 -\tilde{\beta}_0^* {d}_0^\dagger,
	\quad
	\tilde{B}_{n,\lambda} = \sum_{\lambda' = \pm}
	\left[\tilde{\alpha}_{n,\lambda}^{(\lambda')}{b}_{n,\lambda'}
	-\tilde{\beta}_{n,\lambda}^{(\lambda')*}{d}_{n,\lambda'}^{\,\dagger}\right],
	\\
	\tilde{D}_0^\dagger &= \tilde\beta_0 {b}_0 + \tilde\alpha_0^* {d}_0^\dagger,
	\quad
	\tilde{D}_{n,\lambda}^\dagger = \sum_{\lambda'=\pm}
	\left[\tilde{\beta}_{n,\lambda}^{(\lambda')}{b}_{n,\lambda'} 
	+ \tilde{\alpha}_{n,\lambda}^{(\lambda')*} {d}_{n,\lambda'}^{\,\dagger}\right],
\end{align}
in the \FCBnsp, respectively (see App.~\ref{app:details} for definitions and the derivation).
The integer $n$ labels the discrete Landau levels, $n = 0$ corresponding to the lowest Landau level
while $n = 1, 2, \cdots$ refers to the higher Landau levels.
Notice that the higher Landau levels are analogous to the situation without a magnetic field;
the only difference is whether the transverse momentum is quantized or not.
The frequencies in the decomposition are different in the two bases. They are obtained as
\begin{align}
	\Omega_0 &= \sqrt{\Pi_z^2 + m^2},
	\quad
	\Omega_n = \sqrt{\Pi_z^2 + m_T^2},
\end{align}
with $m_T = \sqrt{m^2 + m_B^2}$, $m_B = \sqrt{2 n g |Q B|}$, 
$\Theta_0 = \int^t \dd t \,  \Omega_0$ and $\Theta_n = \int^t \dd t \,  \Omega_n$
in the Hamiltonian basis, and 
\begin{align}
	\Omega_{0,s} = \sqrt{(\Pi_z - s\dot{\theta}_{5+m})^2 + m^2},
	\quad
	\Omega_{n,\lambda} = \sqrt{(\Pi + \lambda \dot{\theta}_{5+m})^2+m^2},
\end{align}
with $\Pi = \sqrt{\Pi_z^2 + m_B^2}$, $s = \text{sign}(QB)$,  $\Theta_{0,s} = \int^t \dd t \,  \Omega_{0,s}$
and ${\Theta}_{n,\lambda} = \int^t \dd t \,  \Omega_{n,\lambda}$ in the \FCBnsp, respectively. 
Finally the creation and annihilation operators satisfy
\begin{align}
	\left\{b_{n_1,\lambda_1}(\vec{p}), b_{n_2,\lambda_2}^\dagger(\vec{p}')\right\}
	&= \left\{d_{n_1, \lambda_1}(\vec{p}), d_{n_2, \lambda_2}^\dagger(\vec{p}')\right\}
	=(2\pi)^2\delta_{\lambda_1 \lambda_2}\delta_{n_1 n_2}\delta^{(2)}(\vec{p}-\vec{p}'),
	\\
	\left\{b_{n_1, \lambda_1}(\vec{p}), d_{n_2, \lambda_2}^\dagger(\vec{p}')\right\}
	&= \left\{b_{n_1, \lambda_1}(\vec{p}), d_{n_2, \lambda_2}(\vec{p}')\right\}
	= 0,
\end{align}
and $B$, $D$, $\tilde{B}$ and $\tilde{D}$ satisfy the same equal-time anti-commutators.
As usual, we assume that there is no external field in the infinite past.
This allows us to uniquely define the vacuum as $b_0 \vert 0 \rangle = b_{n,\lambda} \vert 0 \rangle
= d_0 \vert 0 \rangle = d_{n,\lambda} \vert 0 \rangle = 0$.
The physical quantities are then evaluated as the expectation value of the corresponding 
operators with respect to this vacuum.

Let us now study the Hamiltonian. 
The conjugate momentum follows from Eq.~\eqref{eq:full_action} as
\begin{align}
	\pi_\phi &= \dot{\phi} + \frac{c_5}{f_a}\bar{\psi}\gamma^0 \gamma_5 \psi,
	\quad
	\pi_\psi = i\bar{\psi}\gamma^0,
	\quad
	\pi_{A_i} = -F^{0i} + c_A\frac{\alpha}{\pi f_a}\phi\,\epsilon^{0ijk}F_{jk},
\end{align}
and thus the Hamiltonian density operator is given by
\begin{align}
	\mathcal{H} = \frac{1}{2}\dot{\phi}^2 
	+ V(\phi) +\psi^\dagger\left(i \partial_0 + \dot{\theta}_5 \gamma_5\right)\psi
	+ \frac{1}{2}\left(\vec{E}^2 + \vec{B}^2\right).
\end{align}
where we have assumed that the fermion satisfies the Dirac equation, and substituted our configuration of
the axion and gauge field.
Thus we define the fermionic part of the Hamiltonian as 
\begin{align}
	H_\psi \equiv \frac{1}{2}\int \dd^3 x\left[\psi^\dagger, \left(i \partial_0 + \dot{\theta}_5 \gamma_5\right)\psi\right],
	\label{eq:Hpsi}
\end{align}
where we have anti-symmetrized the operator. 
Note that ${H}_\psi$ contains not only the time derivative
of the fermion but also $\dot{\theta}_5$ explicitly.
In the Hamiltonian basis, by substituting Eq.~\eqref{eq:quantization_Hamiltonian}, we obtain
\begin{align}
	H_\psi &= \frac{1}{2}\int\frac{\dd p_y \dd p_z}{(2\pi)^2}
	\left[\Omega_0 \left(\left[B_0^\dagger, B_0\right] + \left[D_0^\dagger, D_0\right] \right)
	+ \sum_{n,\lambda}\Omega_n \left(\left[B_{n,\lambda}^\dagger, B_{n,\lambda}\right]
	+ \left[D_{n,\lambda}^\dagger, D_{n,\lambda}\right]
	\right) \right].
\end{align}
This shows that the Hamiltonian basis diagonalizes the Hamiltonian even at the intermediate times for a time-dependent background,
as the name already indicated. In particular, $\Omega_0$ and $\Omega_{n}$ are the energy eigenvalues,
which we can then use for regularization without spoiling any symmetry.
On the other hand, in the \FCB, 
the Hamiltonian is expressed by substituting Eq.~\eqref{eq:quantization_derivative} as
\begin{align}
	H_\psi =& \frac{1}{2}\int\frac{\dd p_y \dd p_z}{(2\pi)^2}
	\left\{\frac{\Pi_z^2 + m^2 -s \dot{\theta}_{5+m}\Pi_z}{\Omega_{0,s}}
	\left(\left[\tilde{B}_0^\dagger, \tilde{B}_0\right] + \left[\tilde{D}_0^\dagger, \tilde{D}_0\right] \right)
	- \frac{m \dot{\theta}_{5+m}}{\Omega_{0,s}}
	\left(e^{2i\Theta_{0,s}}\left[\tilde{B}_0^\dagger, \tilde{D}_0^\dagger\right] + (\mathrm{h.c.})\right)
	\right. \nonumber \\
	&\left.
	+ \sum_{n,\lambda}\left[
	\frac{\Pi_B^2 + m^2 +\lambda \dot{\theta}_{5+m}\Pi_B }{\Omega_{n,\lambda}}\left(
	\left[\tilde{B}_{n,\lambda}^\dagger, \tilde{B}_{n,\lambda}\right] 
	+ \left[\tilde{D}_{n,\lambda}^\dagger, \tilde{D}_{n,\lambda}\right]
	\right)
	+ \frac{m \dot{\theta}_{5+m}}{\Omega_{n,\lambda}}
	\left(e^{2i\Theta_{n,\lambda}}\left[\tilde{B}_{n,\lambda}^\dagger, \tilde{D}_{n,\lambda}^\dagger\right] 
	+ (\mathrm{h.c.})\right)
	\right]
	\right\}.
\end{align}
This shows that the \FCB \emph{does not} diagonalize the Hamiltonian.
The off-diagonal parts (the second and fourth terms) originate from the second term in Eq.~\eqref{eq:Hpsi}.
This in particular means that $\Omega_{0,s}$ and $\Omega_{n,\pm}$ are \emph{not} energy eigenvalues.
We may then anticipate that the use of regulator functions based on these frequencies will fail to produce the correct regularized theory.
We will investigate this point in Sec.~\ref{subsec:anomaly}, using the anomaly equation as an explicit example.

%%%%%%%%%%%%%
\subsection{Regularization and anomaly equation}
\label{subsec:anomaly}
%%%%%%%%%%%%%
It is well-known that the chiral rotation is anomalous in the presence of an gauge field.
As a result, the divergence of the axial current $J^\mu_5$ satisfies the anomaly equation:
\begin{align}
	\partial_\mu J^\mu_5 = - \frac{g^2 Q^2}{8\pi^2}F_{\mu\nu}\tilde{F}^{\mu\nu}
	+ 2m\, \bar{\psi} e^{2i\theta_m \gamma_5} i\gamma_5 \psi,
\end{align}
where $J^\mu_5 = \bar{\psi} \gamma^\mu \gamma_5 \psi$.
The anomaly equation is an operator relation and thus must hold in any background field configuration.
This relation provides a non-trivial check of the computation, in particular of the regularization procedure,
since an improper regularization easily destroys the anomaly equation.

As demonstrated in Ref.~\cite{Domcke:2021fee} the regularization based on the Hamiltonian basis
correctly reproduces the anomaly equation, as we briefly review here.
We consider the spatially-averaged version of the anomaly equation.
With our gauge field configuration~\eqref{eq:backgroundB} this is given by
\begin{align}
	\dot{q}_5 = \frac{g^2 Q^2 E B}{2\pi^2} 
	+ 2 m \left\langle  \bar{\psi} e^{2i\theta_m \gamma_5} i\gamma_5 \psi \right\rangle,
	\label{eq:anomaly_spatial_average}
\end{align}
where we have replaced $F_{\mu\nu}$ by the electric and magnetic fields, and
\begin{align}
	q_5 &\equiv \frac{1}{2\mathrm{Vol}\left(\mathbb{R}_3\right)}\int \dd^3x
	\left\langle 
	\left[ {\psi}^\dagger, \gamma_5 \psi\right]
	\right\rangle,
	\quad
	\left\langle \bar{\psi} e^{2i\theta_m \gamma_5} i\gamma_5 \psi \right\rangle
	\equiv \frac{1}{2\mathrm{Vol}\left(\mathbb{R}_3\right)}\int \dd^3x
	\left\langle 
	\left[ \bar{\psi}, i\gamma_5 e^{2i\theta_m \gamma_5} \psi\right]
	\right\rangle.
\end{align}
The lowest Landau level contribution to the chiral charge is given by
\begin{align}
	\left.q_{5}\right\vert_{\mathrm{LLL}} &= \frac{g\abs{QB}}{4\pi^2}\int \dd p_z 
	\left[
	\frac{s{\Pi}_z}{\Omega_0} \left(2\left\lvert \beta_0 \right\rvert^2 - R\right) 
	+ \frac{m}{\Omega_0}\left(\alpha_0 \beta_0^*e^{-2i\Theta_0} + \alpha^*_0 \beta_0 e^{2i\Theta_0} \right)
	\right],
\end{align}
where we have introduced the regulator function $R = R(\Omega_0)$.
In particular, normal ordering corresponds to $R = 0$.
The time derivative is given by
\begin{align}
	\left.\dot{q}_{5}\right\vert_{\mathrm{LLL}} = \frac{g\abs{QB}}{4\pi^2} \int \dd p_z
	\left[
	s\frac{m^2 \dot{{\Pi}}_z}{\Omega_0^3}\left(1 - R \right)
	- s\frac{{\Pi}_z}{\Omega_0}\dot{R} - 2im \left(\alpha_0 \beta_0^* e^{-2i\Theta_0} - \alpha_0^* \beta_0 e^{2i\Theta_0}\right)
	\right],
\end{align}
where we have used the time evolution equation of the Bogoliubov coefficients.
We thus obtain
\begin{align}
	\left.\dot{q}_{5}\right\vert_{\mathrm{LLL}} = \frac{gQB}{4\pi^2} \int \dd p_z
	\left[
	\frac{m^2 \dot{{\Pi}}_z}{\Omega_0^3}\left(1 - R \right)
	- \frac{{\Pi}_z}{\Omega_0}\dot{R}	
	\right]
	+ 2m\left.\left\langle \bar{\psi} e^{2i\theta_m \gamma_5} i\gamma_5 \psi \right\rangle\right\vert_{\mathrm{LLL}}.
\end{align}
The integral does not depend on the explicit form of $R = R(\Omega_0)$,
and we obtain
\begin{align}
	\left.\dot{q}_{5}\right\vert_{\mathrm{LLL}} = \frac{g^2Q^2EB}{2\pi^2} 
	+ 2m \left.\left\langle \bar{\psi} e^{2i\theta_m \gamma_5} i\gamma_5 \psi \right\rangle\right\vert_{\mathrm{LLL}}.
\end{align}
This shows that the anomalous term is supplied solely by the lowest Landau level.
Indeed one can see that the higher Landau level satisfies
\begin{align}
	\left.\dot{q}_{5}\right\vert_{\mathrm{HLL}} &= 
	2m \left.\left\langle \bar{\psi}_n e^{2i\gamma_5 \theta_m}i\gamma_5 \psi_n \right\rangle\right\vert_{\mathrm{HLL}}.
\end{align}
We have thus proven that the anomaly equation~\eqref{eq:anomaly_spatial_average} indeed holds.

We now study the anomaly equation in the \FCBnsp.
We saw that the \FCB does not diagonalize the Hamiltonian.
It is then naturally expected that
the frequencies in this basis, $\Omega_{0,s}$ and $\Omega_{n,\lambda}$, 
are not suitable for regularization as they are not related to the energy eigenvalues.
Indeed, in the following, we will see that the anomaly equation \emph{does not} hold
if one uses $\Omega_{0,s}$ and $\Omega_{n,\lambda}$ as the argument of the regulator function.
This highlights the main point of this section; the regularization is straightforward in the Hamiltonian basis, but 
is more tricky in the \FCB.

If we use $\Omega_{0,s}$ and $\Omega_{n,\lambda}$ for regularization, 
the lowest Landau level contribution may be written as
\begin{align}
	\left.q_5\right\vert_{\mathrm{LLL}} &= 
	\frac{g\abs{QB}}{4\pi^2}\int \dd p_z
	\left[
	\frac{s\Pi_z-\dot{\theta}_{5+m}}{\Omega_{0,s}}
	\left(2\abs{\tilde\beta_0}^2 - \tilde{R}(\Omega_{0,s})\right)
	+ \frac{2m}{\Omega_{0,s}}\mathrm{Re}\left[\tilde\alpha_0^* \tilde\beta_0 e^{2i\Theta_{0,s}}\right] \right],
\end{align}
with a regulator function $\tilde{R}$.
Notice that this includes the normal ordering in the \FCB as a special case for $\tilde{R} = 0$.
By taking the time derivative and using the time evolution equation of the Bogoliubov coefficients,
we obtain
\begin{align}
	\left.\dot{q}_5 \right\vert_{\mathrm{LLL}}
	&= \frac{gQB}{4\pi^2}
	\int \dd p_z\left[
	\frac{m^2(\dot{\Pi}_z - s\ddot{\theta}_{5+m})}{\Omega_{0,s}}
	\left(1-\tilde{R}\right)
	-\frac{{\Pi}_z - s\dot{\theta}_{5+m}}{\Omega_{0,s}}\dot{\tilde{R}}
	\right]
	+2m \left.\left\langle \bar{\psi} e^{2i\theta_m \gamma_5} 
	i\gamma_5 \psi \right\rangle\right\vert_{\mathrm{LLL}}.
\end{align}
The integral again does not depend on the explicit form of $\tilde{R}$ as long as $\tilde{R} = \tilde{R}(\Omega_{0,s})$,
and we obtain
\begin{align}
	\left.\dot{q}_5 \right\vert_{\mathrm{LLL}}
	&= \frac{gQB}{2\pi^2}\left(gQE - s\ddot{\theta}_{5+m}\right)
	+2m \left.\left\langle \bar{\psi} e^{2i\theta_m \gamma_5} 
	i\gamma_5 \psi \right\rangle\right\vert_{\mathrm{LLL}}.
\end{align}
Note the additional term proportional to $\ddot{\theta}_{5+m}$.
One can check (after a somewhat tedious computation) that the contributions from the higher Landau levels
do not cancel this additional term.
Therefore, the regularization based on the frequencies
in the \FCBnsp,  $\Omega_{0,s}$ and $\Omega_{n,\lambda}$, 
\emph{does not} reproduce the anomaly equation.
This is understandable because the \FCB does not diagonalize the Hamiltonian,
and thus the regularization with $\Omega_{0,s}$ and $\Omega_{n,\lambda}$ does not necessarily preserve
all the symmetries that one wants to preserve.
This is in contrast to the Hamiltonian basis;
the Hamiltonian is invariant under all the symmetries by definition,
and thus the regularization with the energy eigenvalues, \emph{i.e.}, the frequencies in the Hamiltonian basis,
automatically preserves all the symmetries.
Needless to say, if we use the same regularization as the Hamiltonian basis,
we correctly reproduce the anomaly equation in the \FCB as well,
as the physics is independent of the basis choice.
The point here is that the regularization is straightforward in the Hamiltonian basis,
but requires special care in the \FCBnsp.
One may prefer to work in the \FCB to study particle production 
(see the end of Sec.~\ref{subsec:axion_schwinger}),
but may prefer to work in the Hamiltonian basis to regularize and renormalize the theory.

Before closing this section, we would like to comment on Ref.~\cite{Adshead:2021ezw}, which computed the fermion contribution to the divergence of the axial current in the \FCB
with dimensional regularization. The authors obtained a term with derivatives acting on the axion that does not vanish
even in the massless fermion limit, on top of the standard Chern-Simons contribution.
From the point of view presented here, this is simply because the regularization scheme employed in \cite{Adshead:2021ezw} does not respect all the symmetries.
Indeed, as the authors noted in the paper, the terms that they got can be canceled by adding local counter terms to the action,
called an irrelevant anomaly, indicating that there is a better regularization scheme. 
This cancellation is not a coincidence or fine tuning, but a reflection of the fact that the contribution from $c_5 + c_m$ vanishes in the massless limit due to the chiral rotation
invariance of the theory.
Therefore, there is a special reason to choose, among others, 
the regularization that preserves this property.
Nevertheless, we agree that there exists a subtlety on higher dimensional operators related to the UV completion,
and we discuss this point in the next section.

%%%%%%%%%%%%%
\section{Transient phenomena and UV sensitivity}
\label{sec:transients}
%%%%%%%%%%%%%

In Sec.~\ref{sec:basis} we demonstrated that the regularization is straightforward in the Hamiltonian basis 
but requires a special care in the \FCBnsp, giving the anomaly equation as an example.
Armed with the proper understanding of the regularization,
we now study another interesting property of the physical quantities at the intermediate times:
exponentially enhanced transient fermion energy density production for finite fermion masses.

As Eq.~\eqref{eq:beta2_axion_assisted} suggests, 
although the axion assisted Schwinger effect lifts the suppression
from the transverse momentum, 
the production rate is still suppressed exponentially by the fermion mass $m$ for $m^2 \gg g\abs{Q}E$.
However, Eq.~\eqref{eq:beta2_axion_assisted} describes the spectrum only at the infinite future,
after turning off the external fields, 
and the physical quantities at the intermediate times do not necessarily feature this suppression factor.
Indeed, as we will see, if the axion velocity is large enough,
the fermion energy density develops a component that is not exponentially suppressed
even for $m^2 \gg g\abs{Q}E$.
With the proper regularization, 
we will check that the same result is obtained in both the Hamiltonian and the {\FCBsnsp}
(as it should), although their appearance is very different;
the transient feature originates from the Bogoliubov coefficients in the Hamiltonian basis 
while it comes from the vacuum contribution in the \FCBnsp.

This phenomena is related to higher dimensional operators in the axion effective 
theory after integrating out fermions.
At the intermediate times, the physical quantities often contain components that are suppressed only by
a power, not an exponential, of the fermion mass in the heavy fermion limit.
These components are in general identified with higher dimensional operators that arise after integrating out the fermion,
such as the Euler-Heisenberg Lagrangian, as described in detail in Refs.~\cite{Banyeres:2018aax,Domcke:2019qmm}.
We will see that the transient effect that we study here also corresponds to higher dimensional operators in the axion effective theory.

Identified with higher dimensional operators, this transient effect is actually degenerate with a choice of the UV theory.
The effective theory with the axion is non-renormalizable, 
and thus there is in general no reason to prohibit adding  higher dimensional operators by hand.
These higher dimensional operators contribute in the same way as the transient effect that we discuss here,
and thus 
giving any physical meaning to the transient effect requires specifying the UV theory.
We believe that it is worth clarifying these points, also to avoid any possible future confusion in literature
(see, \emph{e.g.}\ Refs.~\cite{Kobayashi:2014zza,Stahl:2015gaa,Bavarsad:2016cxh,Hayashinaka:2016qqn,
Hayashinaka:2016dnt,Bavarsad:2017oyv,Hayashinaka:2018amz} 
for a similar confusion on the induced current from the Schwinger effect, 
which turns out to originate merely from the Euler-Heisenberg Lagrangian~\cite{Banyeres:2018aax,Domcke:2019qmm}).
Therefore, we explain all the above points in detail in this section.

%%%%%%%%%%%%%
\subsection{Transient effects and basis choice independence}
\label{subsec:transient}
%%%%%%%%%%%%%
We now study the fermion production for a large axion velocity.
For our purpose in this section, we do not actually need any external electric nor magnetic fields,
and thus we set $E = B = 0$ for simplicity.
In this case there is no particle production at the infinite future,
nevertheless the physical quantities show non-trivial transient behaviors at the intermediate times as we will see.
One can verify that these transient effects also occur if the electromagnetic fields are present.
Without external electromagnetic fields, the equations of the Bogoliubov coefficients are simplified as
\begin{align}
	\begin{pmatrix}
	\dot{\alpha}_1 \\
	\dot{\alpha}_2 \\
	\dot{\beta}_1 \\
	\dot{\beta}_2 
	\end{pmatrix}
	=&
	i\dot{\theta}_{5+m}
	\begin{pmatrix}
	- \frac{m}{m_T}\frac{p_z}{\Omega} & \frac{p_T}{m_T} & \frac{m}{\Omega}e^{2i\Theta} & 0 \\
	\frac{p_T}{m_T} & \frac{m}{m_T}\frac{p_z}{\Omega} & 0 & -\frac{m}{\Omega}e^{2i\Theta} \\
	\frac{m}{\Omega}e^{-2i\Theta} & 0 & \frac{m}{m_T}\frac{p_z}{\Omega} & \frac{p_T}{m_T} \\
	0 & -\frac{m}{\Omega}e^{-2i\Theta} & \frac{p_T}{m_T} & -\frac{m}{m_T}\frac{p_z}{\Omega}
	\end{pmatrix}
	\begin{pmatrix}
	{\alpha}_1 \\
	{\alpha}_2 \\
	{\beta}_1 \\
	{\beta}_2 
	\end{pmatrix},
\end{align}
in the Hamiltonian basis, and
\begin{align}
	\dot{\tilde\alpha}_\lambda &= 
	-\frac{m \ddot{\theta}_{5+m}}{2\Omega_\lambda^2}\exp\left[2i\Theta_\lambda\right] \tilde\beta_\lambda,
	\quad
	\dot{\tilde\beta}_\lambda = 
	\frac{m \ddot{\theta}_{5+m}}{2\Omega_\lambda^2}\exp\left[-2i\Theta_\lambda\right] \tilde{\alpha}_\lambda,
	\label{eq:Bogoliubov_derivative_E=0}
\end{align}
in the \FCBnsp, respectively, where we replace $\Pi_z \to p_z$ and $\Pi \to p$.
We first study the system in the Hamiltonian basis, and then in the \FCBnsp.

%%%%%
\paragraph{Hamiltonian basis.}
%%%%%

%%
\begin{figure}[t]
	\centering
 	\includegraphics[width=0.495\linewidth]{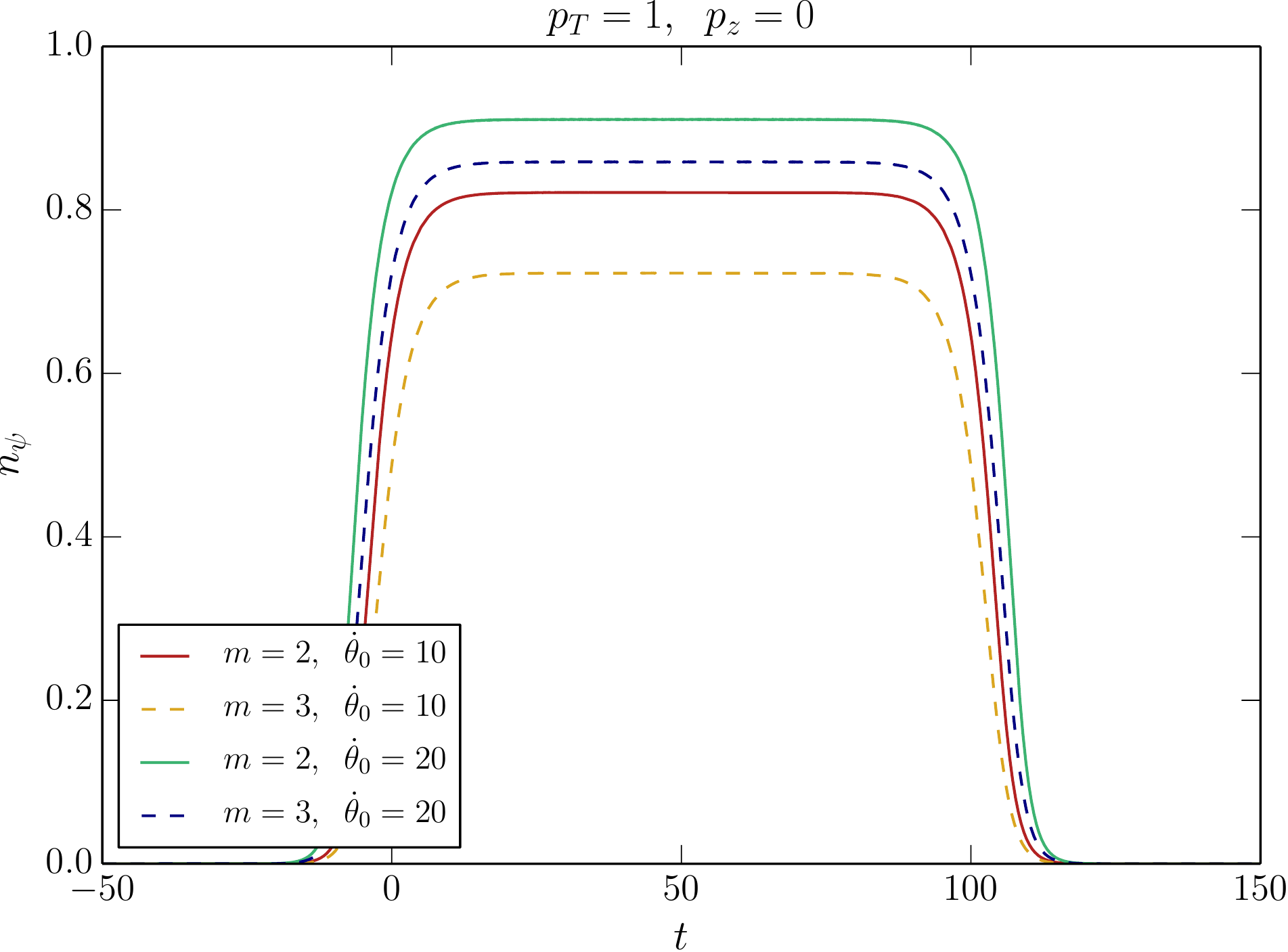}
 	\includegraphics[width=0.495\linewidth]{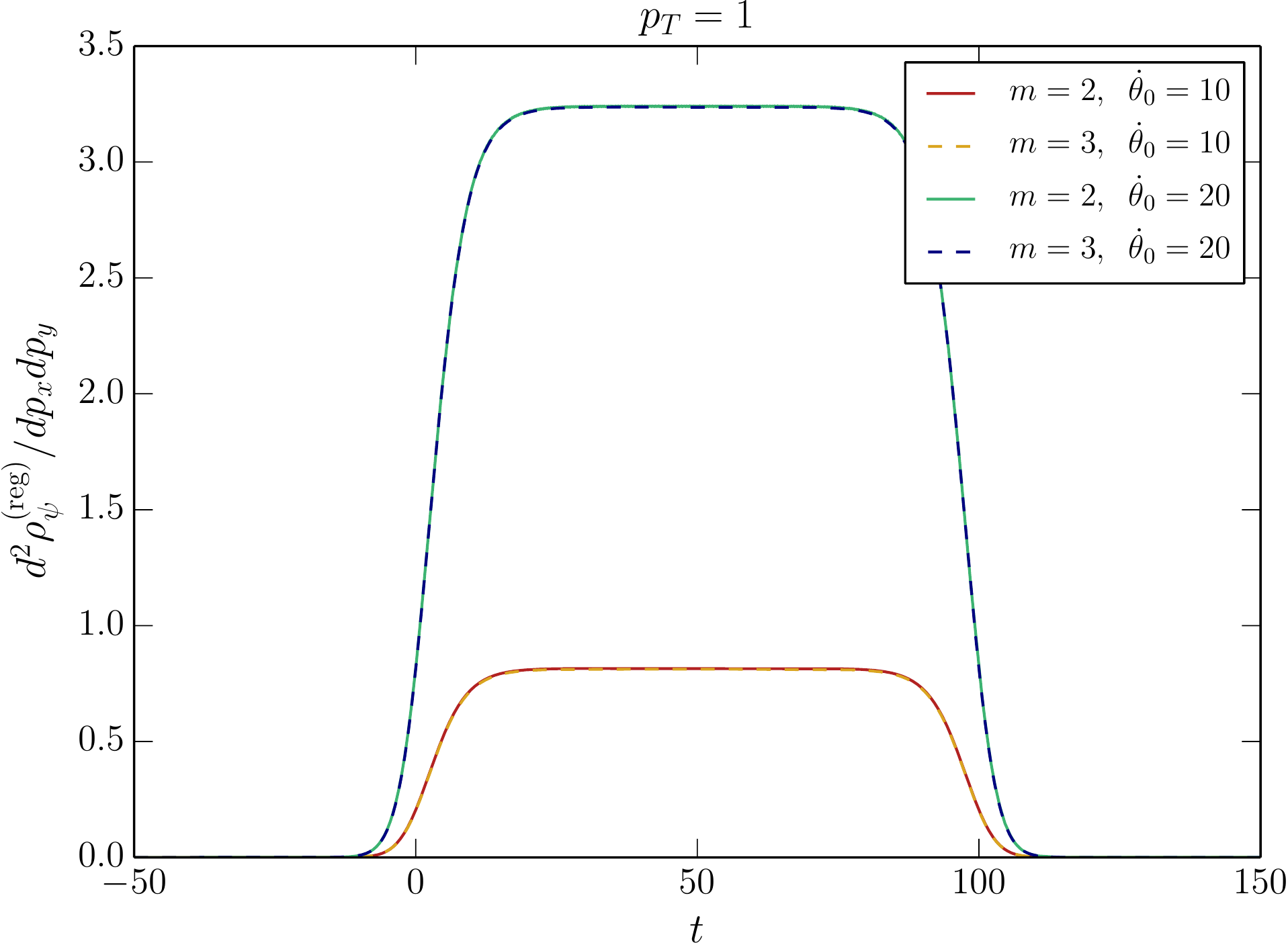}
	\caption{\small
	\emph{Left:} Time evolution of the occupation number $n_\psi$ in the Hamiltonian basis for a large axion velocity $\dot \theta_0$ and vanishing electromagnetic fields.
	Although approaching to zero at the asymptotic future, $n_\psi$ is $\mathcal{O}(1)$ at the intermediate time
	for large $\dot{\theta}_0$.
	We fix $p_z = 0$, and show the results for several different values of $m$ and $\dot{\theta}_0$ in the unit $p_T = 1$.
	\emph{Right:} Time evolution of the regularized (differential) energy density $\dd^2 \rho_\psi^{(\mathrm{reg})}/\dd p_x \dd p_y$ in the Hamiltonian basis with a large axion velocity,
	where we integrate over $p_z$ in the range of $-150$ to $150$.
  This clearly shows that the transient effect in $n_\psi$ is reflected in the physical energy density.}
	\label{fig:transient_Hamiltonian}
\end{figure}

We show our numerical results for $n_\psi = \sum_{\lambda,\lambda'}\vert\beta_{\lambda}^{(\lambda')}\vert^2$ 
as a function of time in the left panel of Fig.~\ref{fig:transient_Hamiltonian}.
We fix $p_z = 0$, $t_\mathrm{min} = 0$, $t_\mathrm{max} = 100$ and $T=\sqrt{50}$, 
and take several different values of $m$ and $\dot{\theta}_0$ in the unit $p_T = 1$.
As one can see, even though the fermion mass is large and there is no external electric field,
$n_\psi$ becomes of $\mathcal{O}(1)$ in the intermediate times.
This is the transient effect that we study in this section.
Although $n_\psi$ itself is not a physical quantity, the same feature appears in the physical quantities such as the energy density.
The fermion energy density in the Hamiltonian basis is expressed as
\begin{align}
	\rho_\psi &\equiv \frac{1}{\mathrm{Vol}(\mathbb{R}_3)} \langle H_\psi \rangle
	= \rho_\psi^{(\mathrm{reg})} + \rho_\psi^\mathrm{(vac)},
\end{align}
where the regularized part $\rho_\psi^{\mathrm{(reg)}}$ and the vacuum part $\rho_\psi^{\mathrm{(vac)}}$ 
are respectively given by
\begin{align}
	\rho_\psi^{\mathrm{(reg)}}
	&=\int \frac{\dd^3 p}{\left(2\pi\right)^3}\,2\Omega
	\sum_{\lambda, \lambda'}\abs{\beta_\lambda^{(\lambda')}}^2,
	\quad
	\rho^{\mathrm{(vac)}}
	=-2\int \frac{\dd^3p}{\left(2\pi\right)^3}\,\Omega.
\label{eq:rhoHB}	
\end{align}
We may employ normal ordering to renormalize the energy density,
corresponding to dropping $\rho_\psi^{\mathrm{(vac)}}$. Recall that the normal ordering
in the Hamiltonian basis indeed correctly reproduces the anomaly equation (see Sec.~\ref{subsec:anomaly}).
In the right panel of Fig.~\ref{fig:transient_Hamiltonian}, 
we plot the time evolution of the differential energy density $\dd^2\rho_\psi^{(\mathrm{reg})}/\dd p_x \dd p_y$
(integrated from $p_z = -150$ to $150$).
We see that the energy density shows the same transient effect as $n_\psi$.

%%%%%
\paragraph{Fermion current basis.}
%%%%%

%%
\begin{figure}[t]
	\centering
 	\includegraphics[width=0.495\linewidth]{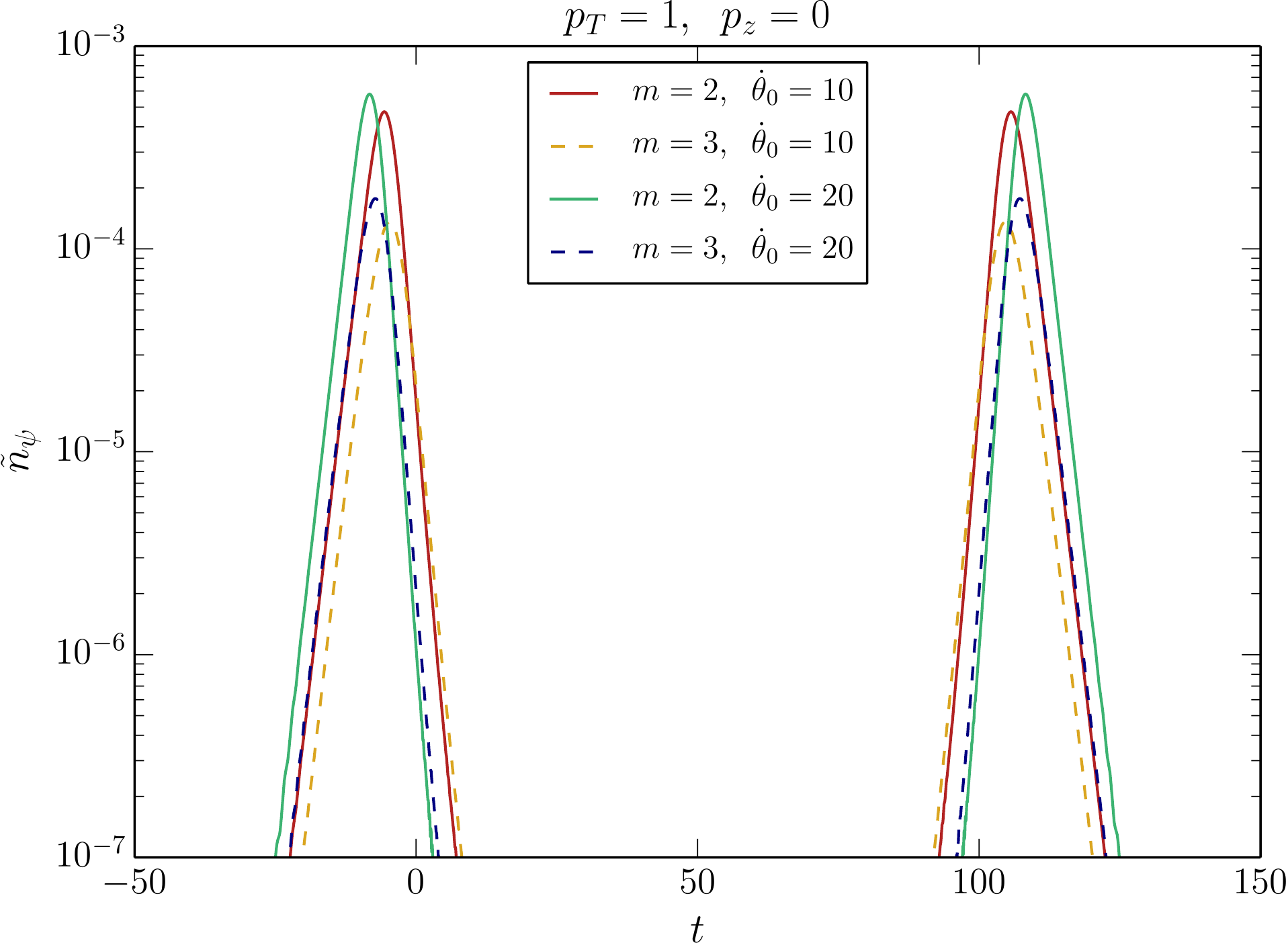}
 	\includegraphics[width=0.495\linewidth]{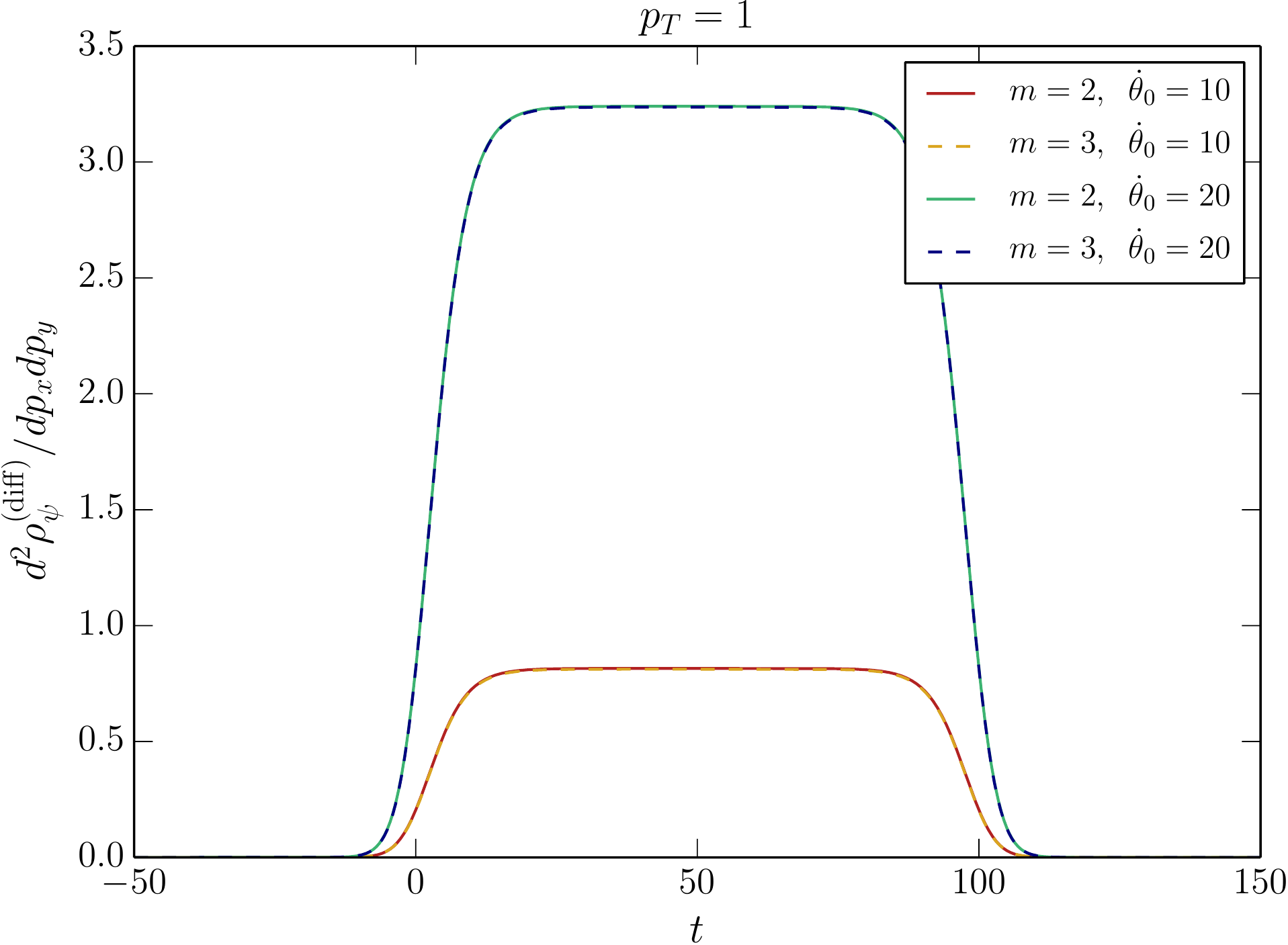}
	\caption{\small 	
	\emph{Left:} Time evolution of the occupation number $\tilde{n}_\psi$ in the \FCB
	with a large axion velocity.
	Even for a large axion velocity, $\tilde{n}_\psi$ is well suppressed and does not show the transient
	effect that shows up in $n_\psi$ in the Hamiltonian basis.
	\emph{Right:} Time evolution of the vacuum contribution to the energy density 
	$\dd^2 \rho_\psi^{(\mathrm{diff})}/\dd p_x \dd p_y$,
	again integrated over $p_z = -150$ to $150$ in the unit $p_T = 1$,
	in the \FCB with a large axion velocity.
	This component features the transient effect, making the total energy density basis-independent.
	The parameters are the same as Fig.~\ref{fig:transient_Hamiltonian}.}
	\label{fig:transient_derivative}
\end{figure}

We now study the same phenomena in the \FCBnsp.
In the left panel of Fig.~\ref{fig:transient_derivative}, 
we plot the time evolution of $\tilde{n}_\psi = \sum_{\lambda,\lambda'}\vert\tilde\beta_{\lambda}^{(\lambda')}\vert^2$,
with the parameters as in Fig.~\ref{fig:transient_Hamiltonian}.
It is clear that $\tilde{n}_\psi$ is well suppressed and does not show this transient effect.\footnote{
	The peaks at $t \sim t_\mathrm{min}$ and $t_\mathrm{max}$ are induced by non-zero $\ddot{\theta}_{5+m}$,
	which depends on the details of turning on and off the axion velocity and is not in the focus of our interest here.
	Note that the spectrum is well below $\mathcal{O}(1)$ even at the peaks.
}
Since $\tilde{n}_\psi$ is not a physical quantity, this by itself does not imply any contradiction in our computation.
In order to investigate the origin of the transient effect in the \FCBnsp,
we take a closer look at the renormalization of the energy density.
In the \FCBnsp, we may decompose the fermion energy density (before the renormalization) as
\begin{align}
	\rho_\psi &= \tilde{\rho}_\psi^{\mathrm{(reg)}}
	+ \rho_\psi^{(\mathrm{diff})} + \rho_{\psi}^{(\mathrm{vac})},
\end{align}
where
\begin{align}
	\tilde{\rho}_\psi^{\mathrm{(reg)}}
	&= \sum_{\lambda, \lambda'}\int \frac{\dd^3p}{(2\pi)^3}
	\left[\frac{p^2 + m^2 + \lambda p \dot{\theta}_{5+m}}{\Omega_\lambda}
	\left(\abs{\tilde\beta_{\lambda}^{(\lambda')}}^2 - \abs{\tilde\alpha_{\lambda}^{(\lambda')}}^2 
	+ \delta_{\lambda\lambda'}\right)
	- 2\dot{\theta}_{5+m}\frac{m}{\Omega_\lambda}
	\mathrm{Re}\left[e^{2i\Theta_{\lambda}} \tilde\alpha_{\lambda}^{(\lambda')*}\tilde\beta_{\lambda}^{(\lambda')}\right]
	\right],
	\\
	\rho_\psi^{\mathrm{(diff)}}
	&= -\sum_{\lambda}\int \frac{\dd^3 p}{(2\pi)^3}
	\left(\frac{p^2 + m^2 + \lambda p \dot{\theta}_{5+m}}{\Omega_\lambda}-\Omega\right).
\end{align}
and $\rho_\psi^{\text{(vac)}}$ is as in Eq.~\eqref{eq:rhoHB}.
Here $\tilde{\rho}_\psi^{\mathrm{(reg)}}$ corresponds to the normal ordering contribution in the \FCBnsp,
while $\rho_\psi^{\mathrm{(diff)}}$ corresponds to the difference of the normal orderings in the Hamiltonian 
and the \FCBnsp.
After renormalizing the energy density by dropping $\rho^{(\mathrm{vac})}_\psi$,
this difference $\rho_\psi^{\mathrm{(diff)}}$ still contributes to the energy density.
In fact it is this term that gives the transient effect in the \FCBnsp,
as one can see in the right panel of Fig.~\ref{fig:transient_derivative}.
Note that, as we saw in Sec.~\ref{subsec:anomaly},
the normal ordering in the \FCB does not correctly reproduce the anomaly equation,
and thus one needs to take $\rho_\psi^{\mathrm{(diff)}}$ into account.
Therefore, the transient effect arises independent of the basis choice,
providing further indication that our regularization and renormalization procedure is consistent.

The momentum integral of $\rho_\psi^{\mathrm{(diff)}}$ can be done exactly, and we obtain
\begin{align}
	\rho_\psi^{\mathrm{(diff)}}
	&= \frac{m^2 \dot{\theta}_{5+m}^2}{4\pi^2}\log\left(\frac{\Lambda^2}{m^2}\right)
	+ \frac{\dot{\theta}_{5+m}^4}{4\pi^2},
	\label{eq:rho_diff_aft_int}
\end{align}
where we have used a hard UV cut-off $\Lambda$ for the momentum integral.
The first term corresponds to the fermion contribution of the axion wavefunction renormalization:
Including the axion contribution to the energy density, we have
\begin{align}
	\rho_\phi + \rho_\psi^{\mathrm{(diff)}}
	&= \frac{1}{2}Z_\phi \dot{\phi}^2 + V(\phi) + \frac{m^2 \dot{\theta}_{5+m}^2}{4\pi^2}\log\left(\frac{\Lambda^2}{m^2}\right)
	+ \frac{\dot{\theta}_{5+m}^4}{4\pi^2},
\end{align}
where we include the axion wavefunction renormalization $Z_\phi$.
By expanding $Z_\phi = 1+\delta_\phi$, we obtain the divergent part 
of the counter term as
\begin{align}
	\left.\delta_\phi\right\vert_\mathrm{div.} = -\frac{(c_5+c_m)^2}{2\pi^2}\frac{m^2}{f_a^2}\log\left(\frac{\Lambda^2}{m^2}\right).
\end{align}
This agrees with the standard Feynman diagrammatic computation 
(see, \emph{e.g.}\ Ref.~\cite{Adshead:2021ezw}).
The second term in~\eqref{eq:rho_diff_aft_int} is an axion dimension-eight effective operator.
It is trivial to check that the second term reproduces the transient effects that we observed numerically.\footnote{
	There are other higher dimensional operators that arise from $\tilde{\rho}_\psi^{(\mathrm{reg})}$,
	but they are subdominant in the present case.
}
Thus, we conclude that the transient effect can be interpreted as an axion higher dimensional operator.
Armed with this understanding, in the next subsection, 
we discuss the resulting UV sensitivity of this transient effect.

%%%%%
\paragraph{Massless limit.}
%%%%%
As an aside,
it is probably appropriate here to comment on the massless limit $m\to 0$,
before discussing the UV sensitivity.
In this limit, all the effects of $\dot{\theta}_{5+m}$ should disappear
as the axion-fermion coupling is unphysical in this limit.

In the Hamiltonian basis, this is trivial to check.
The Bogoliubov coefficient $\beta$ simply vanishes in the limit $m \to 0$ which we have checked numerically.
The situation is more non-trivial in the \FCBnsp, as Eq.~\eqref{eq:rho_diff_aft_int} in the massless limit reads
\begin{align}
	\lim_{m\to 0} \left[\rho_\psi^{\mathrm{(diff)}}\right]
	&= \frac{\dot{\theta}_{5+m}^4}{4\pi^2},
	\label{eq:rho_diff_massless}
\end{align}
which does not vanish.
In the massless limit, there is however another contribution from $\tilde{\rho}_\psi^{(\mathrm{reg})}$,
and we will see in the following that this contribution cancels the contribution of Eq.~\eqref{eq:rho_diff_massless}.
By noting that
\begin{align}
	\lim_{m\to 0} \frac{m}{\Omega_\lambda^2} = \pi \, \delta(p + \lambda \dot{\theta}_{5+m}),
\end{align}
we see that Eq.~\eqref{eq:Bogoliubov_derivative_E=0}
is non-trivial only for $\vert{p+\lambda \dot{\theta}_{5+m}}\vert \lesssim m$ as $m \rightarrow 0$.
In this region, the phase factors in Eq.~\eqref{eq:Bogoliubov_derivative_E=0}
are approximately constant and hence irrelevant for\footnote{
To see this, let us take $\dot{\theta}_{5+m}$ as a time variable, \emph{i.e.},
$\int \dd t \, \Omega_\lambda = \int \dd \dot{\theta}_{5+m}  \, \Omega_\lambda/\ddot{\theta}_{5+m}$.
For $\vert{p+\lambda \dot{\theta}_{5+m}}\vert \lesssim m$, we have $\Omega_\lambda \sim m$  
and $\dot{\theta}_{5+m}$ varies by $\Delta \dot{\theta}_{5+m} \sim m$.
Thus the change in the phase factor during $\vert{p+\lambda \dot{\theta}_{5+m}}\vert \lesssim m$
is roughly $\Delta \dot{\theta}_{5+m}\times \Omega_\lambda/\ddot{\theta}_{5+m} \sim m^2/\ddot{\theta}_{5+m}$.}
\begin{align}
	m^2 \lesssim \vert{\ddot{\theta}_{5+m}}\vert.
	\label{eq:m2vsddth}
\end{align}
Therefore in the limit $m^2 \ll \vert\ddot{\theta}_{5+m}\vert$, we can separate the time domain into two parts,
$\vert{p+\lambda \dot{\theta}_{5+m}}\vert \gtrsim m$ and $\vert{p+\lambda \dot{\theta}_{5+m}}\vert \lesssim m$.
In the former case $\tilde\alpha_\lambda$ and $\tilde\beta_\lambda$ do not evolve in time.
A nontrivial evolution happens only for the latter case, where the equations are given by
\begin{align}
	\dot{\tilde{\alpha}}_\lambda &= 
	-\frac{\ddot{\theta}_{5+m}}{2}\frac{m}{(p +\lambda \dot{\theta}_{5+m})^2 + m^2} \tilde{\beta}_\lambda,
	\quad
	\dot{\tilde{\beta}}_\lambda = \frac{\ddot{\theta}_{5+m}}{2}\frac{m}{(p +\lambda \dot{\theta}_{5+m})^2 + m^2}
	\tilde{\alpha}_\lambda,
\end{align}
where we have dropped the phase factors.
It is convenient to first assume that $\dot{\theta}_{5+m}$ is a monotonic function. 
Later we will see that we can drop this assumption.
If $\dot{\theta}_{5+m}$ is monotonic, we can change the time variable from $t$ to $\dot{\theta}_{5+m}$.
The equations can be then solved analytically. With the initial condition that
$\tilde{\alpha}_\lambda = 1$, $\tilde{\beta}_\lambda = 0$, and $\dot{\theta}_{5+m} = 0$, 
the solutions in the massless limit are
\begin{align}
	\tilde\alpha_\lambda = \Theta(p+\lambda\dot{\theta}_{5+m}),
	\quad
	\tilde\beta_\lambda = \Theta(-\lambda\dot{\theta}_{5+m}-p).
	\label{eq:sol_derivative_massless}
\end{align}
We now come back to the assumption that $\dot{\theta}_{5+m}$ is monotonic.
Even if $\dot{\theta}_{5+m}$ is not monotonic,
we can consider the time domains between $\ddot{\theta}_{5+m} = 0$ separately.
In each domain $\dot{\theta}_{5+m}$ is a good time variable and thus we can repeat the same computation as above.
It is then clear that, for a given mode with fixed $p$, 
the interchange of $\tilde\alpha_\lambda = 1, \tilde\beta_\lambda = 0$ and $\tilde\alpha_\lambda = 0, \tilde\beta_\lambda = 1$ happens 
every time $p + \lambda \dot{\theta}_{5+m}$ crosses zero.
If $p + \lambda \dot{\theta}_{5+m}$ crosses zero an even number of times, then  $\tilde\alpha_\lambda = 1$ and $\tilde\beta_\lambda = 0$ in the end,
while for an odd number of times we have $\tilde\alpha_\lambda = 0$ and $\tilde\beta_\lambda = 1$.
We assume that $\dot{\theta}_{5+m} = 0$ at the beginning, 
then the mode with $p + \lambda \dot{\theta}_{5+m} > 0$ at a given time 
should cross the point $p + \lambda \dot{\theta}_{5+m} = 0$
an even number of times before the given time, while the one with $p +\lambda \dot{\theta}_{5+m} < 0$ should cross the point odd times,
independently of the detailed time evolution of $\dot{\theta}_{5+m}$.
Thus we conclude that Eq.~\eqref{eq:sol_derivative_massless} is correct
independent of the time dependence of $\dot{\theta}_{5+m}$ in the massless limit.
We have also checked that the numerical solution agrees well with Eq.~\eqref{eq:sol_derivative_massless} 
in the massless limit.
Assuming $\dot{\theta}_{5+m} > 0$ for definiteness, we obtain
\begin{align}
	\lim_{m\to 0}\left[\tilde{\rho}_\psi^{(\mathrm{reg})}\right]
	&= 2\int \frac{\dd^3 p}{(2\pi)^3} \frac{p-\dot{\theta}_{5+m}}{\lvert p-\dot{\theta}_{5+m}\rvert}p\,\Theta(\dot{\theta}_{5+m} - p)
	= -\frac{\dot{\theta}_{5+m}^4}{4\pi^2}.
\end{align}
We thus see that 
\begin{align}
	\lim_{m\to 0}\left[\tilde{\rho}_\psi^{(\mathrm{reg})} + {\rho}_\psi^{(\mathrm{diff})}\right]
	= 0,
\end{align}
so that the total energy density vanishes in the massless limit in the \FCB as well, 
providing further evidence for the consistency of our regularization and renormalization.
As soon as the fermion mass becomes finite, the contribution from $\tilde{\rho}_\psi^{(\mathrm{reg})}$
is well suppressed and ${\rho}_\psi^{(\mathrm{diff})}$ dominates,
as we saw in Fig.~\ref{fig:transient_derivative}.
Thus, contrary to Ref.~\cite{Adshead:2021ezw}, we find the expected decoupling of the fermions in the massless limit without the need to introduce any counter terms (see also discussion below).

%%%%%%%%%%%%%
\subsection{Higher dimensional operator and UV sensitivity}
\label{subsec:higher_dim_op}
%%%%%%%%%%%%%
In the previous subsection, we have seen that the transient effect in the fermion energy density can be identified with a higher dimensional operator.
In the \FCBnsp, this arises due to the difference of the vacuum contribution with respect to the normal ordering prescription in the Hamiltonian basis and takes the form
\begin{align}
	\rho_\psi^{\mathrm{(diff)}}
	&= \frac{m^2 \dot{\theta}_{5+m}^2}{4\pi^2}\log\left(\frac{\Lambda^2}{m^2}\right)
	+ \frac{\dot{\theta}_{5+m}^4}{4\pi^2}.
\end{align}
Identifying the second term with a higher dimensional operator,
this is actually degenerate with the choice of the UV theory.
The low energy effective field theory describing the axion is non-renormalizable,
and thus 
there is a priori no constraint to adding higher dimensional operators by hand, on top of the transient effect discussed here.
Note that the situation is different for the Euler-Heisenberg Lagrangian.
In the case of the Euler-Heisenberg Lagrangian, the theory is renormalizable before integrating out a heavy particle,\footnote{
	Strictly speaking there exists the Landau pole which suggests a new scale,
	but we ignore this subtlety here.
}
and it makes sense to set all the higher dimensional operators to vanish before integrating out the heavy particle. 
One can then uniquely determine the coefficients
of the higher dimensional operators after integrating out the heavy particle.

Thus, our conclusion is that a computation of transient effects in the axion effective theory requires the specification of the UV theory.
In the limit $\dot \theta_{5+m} \ll f_a$, our computation of the transient effects within the axion effective field theory is valid. In the opposite case, the axion
effective field theory expansion is invalid.
One then must go back to the original UV theory and study the dynamics within the UV theory.

Before concluding, let us briefly comment on other regularization schemes. In particular, in the \FCBnsp, we can evaluate the axion effective action 
\begin{align}
	\Gamma[\phi] = -i \mathrm{Tr}\log\left[i\slashed{\partial} - m 
	+ \partial_\mu \theta_{5+m} \gamma^\mu\gamma_5\right],
\end{align}
by expanding in powers of $\dot \theta_{5+m}/m$ and evaluating the resulting terms using dimensional regularization. Treating the $\gamma_5$ matrices using the t'Hooft-Veltman prescription~\cite{tHooft:1972tcz}, we reproduce the first term in Eq.~\eqref{eq:rho_diff_aft_int} but obtain a vanishing coefficient for the term proportional to $\dot \theta_{5+m}^4$, in contrast to the second term in Eq.~\eqref{eq:rho_diff_aft_int}. 
This may also be related to somewhat surprising results in the literature, such as the non-restoration of the axial symmetry in the massless fermion limit found in~\cite{Adshead:2021ezw} using dimensional regularization. In fact, the latter result demonstrates that this regularization scheme does not respect the chiral symmetry of the theory in the massless limit (adding to the list of difficulties of dimensional regularization for chiral theories~\cite{tHooft:1972tcz}), thus violating a key requirement of a `good' regularization scheme.

%%%%%%%%%%%%%
\section{Conclusions}
\label{sec:conclusions}
%%%%%%%%%%%%%
In this paper, we have studied fermion production in the presence of a background axion field with non-vanishing velocity and/or electromagnetic fields.
We have included general dimension-five axion-fermion couplings and thus considered the Dirac equation of the form
\begin{align}
	\left[i \slashed{D} - m e^{2i\theta_m \gamma_5} + \partial_\mu \theta_5
	\gamma^\mu \gamma_5\right] \psi = 0,
\end{align}
where $\theta_i = c_i \phi/f_a$ with $\phi$ being the axion.
While the spectrum of the produced fermion at the infinite future, after turning off the external fields,
is studied in detail in~\cite{Domcke:2021fee},
here we have put more focus on physical quantities at intermediate times, obtained before turning off the external fields.
These physical quantities at the intermediate times are of great phenomenological interest.
For instance, in the context of axion inflation, 
the fermion production backreacts to the axion-gauge field system in the form of physical quantities 
such as the induced current, which in turn determines the production of gauge fields, 
gravitational waves and primordial density fluctuations.

In general, physical quantities are divergent in the presence of the external fields,
and thus a proper understanding of regularization and renormalization is necessary 
to evaluate the physical quantities at the intermediate times.
In particular, a proper regularization scheme must preserve the symmetries of the theory as much as possible,
such as the chiral rotation invariance as reflected in the Adler-Bell-Jackiw anomaly equation,
and it can be highly non-trivial to find such a proper regularization scheme 
(see \emph{e.g.}~\cite{Adshead:2021ezw} for a related discussion).
The main observation of this paper is that the theory is most straightforwardly regularized and renormalized
in the $c_5 = 0$ basis (or the Hamiltonian basis).
A chiral rotation $\psi \to e^{ic\phi/f_a \gamma_5}\psi$ shifts the couplings as 
$c_5 \to c_5 - c$ and $c_m \to c_m + c$,
and a different choice of $c$ corresponds to a different choice of the fermion basis that one works with.
This rotation is merely a redundancy of the theory and the final result should not be affected by this choice.
Nevertheless the Hamiltonian basis is most preferable for regularizing and renormalizing the theory,
as the fermionic part of the Hamiltonian is diagonalized only in the Hamiltonian basis 
and not in other bases including the $c_m = 0$ basis (or the \FCBnsp)
employed in~\cite{Kitamoto:2021wzl}.
We have supported our observation
by showing that the regularization using a regulator function depending on the frequencies in the Hamiltonian basis
(corresponding to the energy eigenvalues) correctly reproduces the anomaly equation, 
while the regularization using the frequencies appearing in the \FCB does not.

With the proper understanding of the regularization and the renormalization, we have studied 
 transient effects in the fermion energy density with a large axion velocity.
We have seen that, in the presence of a large axion velocity, the fermion energy density $\rho_\psi$ 
at intermediate times is no longer exponentially suppressed by the fermion mass, but is significantly enhanced even in the absence of external electromagnetic fields and even for constant axion velocity.
This should be compared with the usual formula for the Schwinger effect,
$\rho_\psi \propto \exp(-\pi m^2/g\abs{Q}E)$,
which indicates an exponential suppression of the energy density at the infinite future.
We have confirmed that the same result follows in both the Hamiltonian and {\FCBsnsp}
although its appearance is very different;
the transient effect originates from the Bogoliubov coefficients in the Hamiltonian basis,
while it stems from a vacuum contribution in the \FCBnsp.
The latter expression allows us to identify the transient effect with an axion higher dimensional operator.
We stress that although this operator is degenerate with the choice of UV completion, 
one can study the transient effect using our prescription for regularization and renormalization
once one specifies the UV completion of the theory.
We did not discuss in any detail phenomenological applications of these transient effects. One obvious obstacle is that a significant enhancement of the fermion energy density, as observed in Fig.~\ref{fig:transient_Hamiltonian}, requires a very large axion velocity. This is not only difficult to achieve in realistic models, but also per se indicates a breakdown of the axion effective theory. A phenomenological analysis of this transient effect should thus be conducted in the framework of the original Peccei-Quinn theory, whenever it is relevant.
On the other hand, our results also show that in the limit of small axion velocities, $\dot \theta_{5+m} \ll f_a$, transient effects can be reliably computed in the axion effective field theory.

%%%%%%%%%%%%%
\paragraph{Acknowledgements}
It is a pleasure to thank Peter Adshead,  Hiroyuki Kitamoto and  Masaki Yamada for  helpful discussions related to this project.
Y.E.\ is supported in part by U.S.\ Department of Energy Grant No.~de-sc0011842.
K.M.\ is supported by MEXT Leading Initiative for Excellent Young Researchers Grant No.\ JPMXS0320200430.
%%%%%%%%%%%%%

%%%%%%%%%%%%%
%%%%%%%%%%%%%
\appendix
%%%%%%%%%%%%%
%%%%%%%%%%%%%

%%%%%%%%%%%%%
\newpage
\section{Notation and conventions}
\label{app:notation}
%%%%%%%%%%%%%
Here we summarize our notation and conventions in this paper.
We take the mostly minus convention and define the anti-commutator of the gamma matrices as
\begin{align}
	\left\{\gamma^{\mu}, \gamma^{\nu}\right\} &= \eta^{\mu\nu},
	\quad
	\eta_{\mu\nu} = \mathrm{diag}(1,-1,-1,-1).
\end{align}
It then follows that $\gamma^0$ is hermitian while $\gamma^i$ is anti-hermitian.
Unless specified otherwise, we work with the Weyl representation, given by
\begin{align}
	\gamma^0 = \begin{pmatrix}
	0 & 1 \\
	1 & 0
	\end{pmatrix},
	\quad
	\gamma^{i} = \begin{pmatrix}
	0 & \sigma_i \\
	-\sigma_i & 0
	\end{pmatrix},
\end{align}
where $\sigma_i$ denotes the Pauli matrices, given by
\begin{align}
	\sigma_1 = \begin{pmatrix}
	0 & 1 \\
	1 & 0
	\end{pmatrix},
	\quad
	\sigma_2 = \begin{pmatrix}
	0 & -i \\
	i & 0
	\end{pmatrix},
	\quad
	\sigma_3 = \begin{pmatrix}
	1 & 0 \\
	0 & -1
	\end{pmatrix}.
\end{align}
We define $\gamma_5$ as
\begin{align}
	\gamma_5 = -\frac{i}{4!}\epsilon^{\mu\nu\rho\sigma}\gamma_\mu \gamma_\nu \gamma_\rho \gamma_\sigma
	= i \gamma^0 \gamma^1 \gamma^2 \gamma^3,
\end{align}
where $\epsilon^{\mu\nu\rho\sigma}$ is the totally anti-symmetric tensor with $\epsilon^{0123} = -\epsilon_{0123} = 1$.
In the Weyl representation $\gamma_5$ is given by
\begin{align}
	\gamma_5 &= \begin{pmatrix}
	-1 & 0 \\
	0 & 1
	\end{pmatrix}.
\end{align}
The covariant derivative is given by
\begin{align}
	i D_{\mu} = i \partial_\mu - g QA_{\mu}.
\end{align}
The field strength is defined as $F_{\mu\nu} = \partial_\mu A_\nu - \partial_\nu A_\mu$, and its dual as
\begin{align}
	\tilde{F}_{\mu\nu} = \frac{\epsilon_{\mu\nu\rho\sigma}}{2}F^{\rho\sigma}.
\end{align}
%%

%%%%%%%%%%%%%
\section{Technical details on solutions of Dirac equation}
\label{app:details}
%%%%%%%%%%%%%
In this appendix we summarize the technical details omitted in the main text.
In particular, we explain in detail the solutions of the Dirac equation in the Hamiltonian 
and the \FCBsnsp.

%%%%%
\subsection{Dirac equation with electric field}
%%%%%
We first consider the case with only the electric field and the axion.
The Dirac equation is given by
\begin{align}
	\left[i \slashed{D} - m e^{2i\theta_m \gamma_5} + \partial_\mu \theta_5
	\gamma^\mu \gamma_5\right] \psi = 0,
\end{align}
and we take the background fields as
\begin{align}
	A^{\mu} = \left(0, 0, 0, A_z(t)\right),
	\qquad
	\phi = \phi(t).
\end{align}
Since the background fields do not depend on the space coordinate,
we can move to Fourier space as
\begin{align}
	\psi(x,t) &= \int\frac{\dd^3 p}{(2\pi)^3}e^{i\vec{p}\cdot \vec{x}} \psi(p, t),
\end{align}
where $\vec{p} = (p_x, p_y, p_z)$.
The Dirac equation then reads
\begin{align}
	0 &= \begin{pmatrix} i\partial_0 + \Pi_z - \dot{\theta}_5 & p_x - ip_y & -me^{2i\theta_m} & 0 \\
	p_x + ip_y & i\partial_0 - \Pi_z - \dot{\theta}_5 & 0 & -me^{2i\theta_m} \\
	-me^{-2i\theta_m} & 0 & i\partial_0 - \Pi_z + \dot{\theta}_5 & -(p_x - ip_y) \\
	0 & -me^{-2i\theta_m} & -(p_x + ip_y) & i\partial_0 + \Pi_z + \dot{\theta}_5
	\end{pmatrix}
	\psi.
\end{align}
We solve this equation in the Hamiltonian basis $c_5 = 0$ and the \FCB $c_m = 0$.

In the Hamiltonian basis, we eliminate $c_5$ by $\psi \to e^{i\theta_5 \gamma_5} \psi$.
The solutions of the Dirac equation for constant $\phi$ and $A_z$ are then given by
\begin{align}
	u_1 &= \frac{e^{-i\gamma_5 \theta_{5+m}}}{N}
	\begin{pmatrix}
	-(p_x - ip_y) m_T \\ \left(\Omega + \Pi_z\right)\left(m + m_T\right) \\
	(p_x - ip_y) \left(\Omega + \Pi_z\right) \\ m_T\left(m + m_T\right)
	\end{pmatrix},
	\quad
	u_2 = \frac{e^{-i\gamma_5 \theta_{5+m}}e^{-i\varphi_p}}{N}
	\begin{pmatrix}
	m_T\left(m + m_T\right) \\  -(p_x + ip_y)\left(\Omega + \Pi_z\right) \\
	 \left(\Omega + \Pi_z\right)\left(m+m_T\right) \\ (p_x + ip_y) m_T
	\end{pmatrix}, \\
	v_1 &= \frac{e^{-i\gamma_5 \theta_{5+m}}}{N}
	\begin{pmatrix}
	-(p_x - ip_y) \left(\Omega + \Pi_z\right) \\  -m_T\left(m + m_T\right) \\
	-(p_x - ip_y)m_T \\ \left(\Omega + \Pi_z\right)\left(m + m_T\right)
	\end{pmatrix},
	\quad
	v_2 = \frac{e^{-i\gamma_5 \theta_{5+m}}e^{-i\varphi_p}}{N}
	\begin{pmatrix}
	 \left(\Omega + \Pi_z\right)\left(m + m_T\right)  \\  (p_x + ip_y) m_T \\
	 -m_T\left(m + m_T\right) \\ (p_x + ip_y)\left(\Omega + \Pi_z\right)
	\end{pmatrix},
\end{align}
with
\begin{align}
	N = 2\sqrt{\Omega\left(\Omega + \Pi_z\right)\left(m + m_T\right)m_T},
	\quad
	p_x - ip_y = p_T e^{-i\varphi_p},
	\quad
	m_T = \sqrt{p_T^2 + m^2}.
\end{align}
Once we turn on the time dependence of $A_z$ and $\phi$, the Bogoliubov coefficients evolve as
\begin{align}
	\begin{pmatrix}
	\dot{\alpha}_1 \\
	\dot{\alpha}_2 \\
	\dot{\beta}_1 \\
	\dot{\beta}_2 
	\end{pmatrix}
	=&
	\left[
	i\dot{\theta}_{5+m}
	\begin{pmatrix}
	- \frac{m}{m_T}\frac{\Pi_z}{\Omega} & \frac{p_T}{m_T} & \frac{m}{\Omega}e^{2i\Theta} & 0 \\
	\frac{p_T}{m_T} & \frac{m}{m_T}\frac{\Pi_z}{\Omega} & 0 & -\frac{m}{\Omega}e^{2i\Theta} \\
	\frac{m}{\Omega}e^{-2i\Theta} & 0 & \frac{m}{m_T}\frac{\Pi_z}{\Omega} & \frac{p_T}{m_T} \\
	0 & -\frac{m}{\Omega}e^{-2i\Theta} & \frac{p_T}{m_T} & -\frac{m}{m_T}\frac{\Pi_z}{\Omega}
	\end{pmatrix}
	+ \frac{m_T \dot{\Pi}_z}{2\Omega^2}
	\begin{pmatrix}
	0 & 0 & -e^{2i\Theta} & 0 \\
	0 & 0 & 0 & -e^{2i\Theta} \\
	e^{-2i\Theta} & 0 & 0 & 0 \\
	0 & e^{-2i\Theta} & 0 & 0
	\end{pmatrix}
	\right]
	\begin{pmatrix}
	{\alpha}_1 \\
	{\alpha}_2 \\
	{\beta}_1 \\
	{\beta}_2 
	\end{pmatrix},
	\label{eq:alphabeta}
\end{align}
where $\Theta = \int^t \dd t \, \Omega$.
We solve these equations with two different initial conditions,
\begin{align}
	\alpha_{1}^{(1)} &= 1,
	\quad
	\alpha_{2}^{(1)} = \beta_{1}^{(1)} = \beta_{2}^{(1)} = 0,
	\qquad\mathrm{and}\qquad
	\alpha_{2}^{(2)} = 1,
	\quad
	\alpha_{1}^{(2)} = \beta_{1}^{(2)} = \beta_{2}^{(2)} = 0.
\end{align}
The former corresponds to the mode which is initially $u_1$,
while the latter is the one which is initially $u_2$.
The Bogoliubov coefficients then describe the mixing of these two positive frequency modes with the others
at a later time.
In terms of the original field in the coordinate space, the fermion is thus quantized as
\begin{align}
	\psi &= \int\frac{\dd^3 p}{(2\pi)^3}e^{i\vec{p}\cdot\vec{x}}
	e^{i\theta_5 \gamma_5}\sum_{\lambda, \lambda'=1,2}\left[
	u_\lambda e^{-i\Theta}
	\left(\alpha_\lambda^{(\lambda')}b_{\lambda'} 
	- (-)^{\lambda+\lambda'}{\beta_{\lambda}^{(\lambda')}}^{*} d_{\lambda'}^\dagger\right) 
	+ v_\lambda e^{+i\Theta}
	\left(\beta_\lambda^{(\lambda')}b_{\lambda'} + (-)^{\lambda+\lambda'}
	{\alpha_\lambda^{(\lambda')}}^* d_{\lambda'}^\dagger\right)\right].
\end{align}

In the \FCBnsp, we instead eliminate $c_m$ by $\psi \to e^{-i\theta_m\gamma_5}\psi$.
In this case we may define the solutions of the Dirac equation for constant $A_z$ and $\dot{\phi}$ as
\begin{align}
	\tilde{u}_+ &= \frac{1}{N_+}
	\begin{pmatrix}
	-(p_x - ip_y)(\Omega_+ + \Pi + \dot{\theta}_{5+m}) \\
	(\Pi+\Pi_z)(\Omega_+ + \Pi + \dot{\theta}_{5+m}) \\
	-(p_x - ip_y)m \\
	(\Pi+\Pi_z) m
	\end{pmatrix},
	\quad
	\tilde{u}_- = \frac{1}{N_-}
	\begin{pmatrix}
	(p_x - ip_y)m \\
	(\Pi-\Pi_z) m \\
	(p_x - ip_y)(\Omega_- + \Pi - \dot{\theta}_{5+m}) \\
	(\Pi-\Pi_z)(\Omega_- + \Pi - \dot{\theta}_{5+m})
	\end{pmatrix},
	\\
	\tilde{v}_+ &= \frac{1}{N_+}
	\begin{pmatrix}
	(p_x - ip_y)m \\
	-(\Pi+\Pi_z) m \\
	-(p_x - ip_y)(\Omega_+ + \Pi + \dot{\theta}_{5+m}) \\
	(\Pi+\Pi_z)(\Omega_+ + \Pi + \dot{\theta}_{5+m})
	\end{pmatrix},
	\quad
	\tilde{v}_- = \frac{1}{N_-}
	\begin{pmatrix}
	-(p_x - ip_y)(\Omega_- + \Pi - \dot{\theta}_{5+m}) \\
	-(\Pi-\Pi_z)(\Omega_- + \Pi - \dot{\theta}_{5+m}) \\
	(p_x - ip_y)m \\
	(\Pi-\Pi_z) m
	\end{pmatrix},
\end{align}
with
\begin{align}
	\Pi &= \sqrt{\Pi_z^2 + p_x^2 + p_y^2},
	\quad
	\Omega_\pm = \sqrt{(\Pi \pm \dot{\theta}_{5+m})^2 + m^2},
	\quad
	N_\pm = 2\sqrt{\Pi(\Pi\pm \Pi_z) \Omega_\pm (\Omega_\pm + \Pi \pm \dot{\theta}_{5+m})},
\end{align}
and $\Theta_\pm = \int^t \Omega_\pm \dd t$.
Once we turn on the time dependence of the background fields,
the Bogoliubov coefficients satisfy
\begin{align}
	\begin{pmatrix}
	\dot{\tilde\alpha}_+ \\
	\dot{\tilde\alpha}_- \\
	\dot{\tilde\beta}_+ \\
	\dot{\tilde\beta}_-
	\end{pmatrix}
	&= \begin{pmatrix}
	0 & Ae^{i(\Theta_+ - \Theta_-)} & -B_+ e^{2i\Theta_+} & -C e^{i(\Theta_+ + \Theta_-)} \\
	-A e^{-i(\Theta_+ - \Theta_-)} & 0 & -Ce^{i(\Theta_+ + \Theta_-)} & -B_-e^{2i\Theta_-} \\
	B_+ e^{-2i\Theta_+} & C e^{-i(\Theta_+ + \Theta_-)}& 0 & A e^{-i(\Theta_+ - \Theta_-)}  \\
	C e^{-i(\Theta_+ + \Theta_-)} & B_- e^{-2i\Theta_-} & -A e^{i(\Theta_+ - \Theta_-)}  & 0
	\end{pmatrix}
	\begin{pmatrix}
	\tilde{\alpha}_+ \\
	\tilde{\alpha}_- \\
	\tilde{\beta}_+ \\
	\tilde{\beta}_-
	\end{pmatrix},
\end{align}
where
\begin{align}
	A &= \frac{gQE\,p_T m }{4\Pi^2 \sqrt{\Omega_+ \Omega_-}}
	\left[\frac{\Omega_+ + \Omega_- + 2\Pi}{\sqrt{\left(\Omega_+ + \Pi + \dot{\theta}_{5+m}\right)
	\left(\Omega_- + \Pi - \dot{\theta}_{5+m}\right)}}\right], \\
	B_{\pm} &= \pm \frac{m}{2\Omega_{\pm}^2}\left(\frac{\Pi_z}{\Pi} gQE \pm \ddot{\theta}_{5+m}\right),
	\quad
	C = \frac{gQE\,p_T 
	\left[ \left(\Omega_+ + \Pi + \dot{\theta}_{5+m}\right)\left(\Omega_- + \Pi - \dot{\theta}_{5+m}\right) - m^2\right]}
	{4\Pi^2 \sqrt{\Omega_+\Omega_- \left(\Omega_+ + \Pi + \dot{\theta}_{5+m}\right)
	\left(\Omega_- + \Pi - \dot{\theta}_{5+m}\right)}}.
\end{align}
We solve these equations with two different initial conditions,
\begin{align}
	\tilde\alpha_{+}^{(+)} &= 1,
	\quad
	\tilde\alpha_{-}^{(+)} = \tilde\beta_{+}^{(+)} = \tilde\beta_{-}^{(+)} = 0,
	\qquad\mathrm{and}\qquad
	\tilde\alpha_{-}^{(-)} = 1,
	\quad
	\tilde\alpha_{+}^{(-)} =\tilde \beta_{+}^{(-)} = \tilde\beta_{-}^{(-)} = 0.
\end{align}
In this basis the Dirac field is quantized as
\begin{align}
	\psi = \int \frac{\dd^3 p}{(2\pi)^3} e^{i\vec{p}\cdot \vec{x}} 
	e^{-i\theta_m\gamma_5}\sum_{\lambda \lambda' = \pm}
	\left[\tilde{u}_\lambda e^{-i\Theta_\lambda}\left(\tilde{\alpha}_\lambda^{(\lambda')}{b}_{\lambda'}
	-\tilde{\beta}_\lambda^{(\lambda')*}{d}_{\lambda'}^{\,\dagger}\right)
	+ \tilde{v}_\lambda e^{i\Theta_\lambda}\left(\tilde{\beta}_\lambda^{(\lambda')}{b}_{\lambda'} 
	+ \tilde{\alpha}_\lambda^{(\lambda')*} {d}_{\lambda'}^{\,\dagger}\right)
	\right],
\end{align}
where we use that $(\tilde\alpha_+, \tilde\alpha_-, \tilde\beta_+, \tilde\beta_-)^{T}$
and $(-\tilde\beta_+, -\tilde\beta_-, \tilde\alpha_+, \tilde\alpha_-)^\dagger$ satisfy the same equation.
We do not put tildes on the creation and annihilation operators here. Since these are the operators at the infinite past
where we assume there is no external field, 
these are equivalent to those in the Hamiltonian basis (up to linear transformation in the case of degenerate eigenvalues).

The creation and annihilation operators at the infinite past are quantized as
\begin{align}
	\left\{b_{\lambda_1}(\vec{p}), b_{\lambda_2}^\dagger(\vec{p}')\right\}
	&= \left\{d_{\lambda_1}(\vec{p}), d_{\lambda_2}^\dagger(\vec{p}')\right\}
	=(2\pi)^3\delta_{\lambda_1 \lambda_2}\delta^{(3)}(\vec{p}-\vec{p}'),
	\\
	\left\{b_{\lambda_1}(\vec{p}), d_{\lambda_2}^\dagger(\vec{p}')\right\}
	&= \left\{b_{\lambda_1}(\vec{p}), d_{\lambda_2}(\vec{p}')\right\}
	= 0,
\end{align}
It then follows from the time evolution of the Bogoliubov coefficients that 
$B_\lambda$, $D_\lambda$, $\tilde{B}_\lambda$ and $\tilde{D}_\lambda$
satisfy the same equal time anti-commutation relations.
The proof is rather lengthy, and we refer interested readers to App.~D of~\cite{Domcke:2021fee},
where it is shown explicitly that the creation and annihillation operators in the Hamiltonian basis,
$B_\lambda$ and $D_\lambda$, satisfy the same equal-time anti-commutators as $b_\lambda$ and $d_\lambda$.
One can then show that $\tilde{B}_\lambda$ and $\tilde{D}_\lambda$ satisfy the same equal-time
anti-commutators by expressing them in terms of $B_\lambda$ and $D_\lambda$, with the help of the completeness condition
\begin{align}
	\sum_{\lambda}\left[u_\lambda u^\dagger_\lambda + v_\lambda v^\dagger_\lambda\right] = \mathbbm{1}_4.
\end{align}
%%

%%%%%
\subsection{Dirac equation with electric and magnetic field}
%%%%%
We now include the magnetic field.
We take the background fields as
\begin{align}
	A^{\mu} = \left(0, 0, B x, A_z\right),
	\qquad
	\phi = \phi(t).
\end{align}
The background fields do not explicitly depend on $y$ and $z$, and thus we perform the Fourier transformation as
\begin{align}
	\psi = \int \frac{\dd p_y \dd p_z}{(2\pi)^2}e^{i(p_y y + p_z z)} \psi(x, p_y, p_z, t).
\end{align}
By using the explicit forms of the gamma matrices, we obtain
\begin{align}
	0 = &\left[i \mathbbm{1}_4 \partial_0 + i \sqrt{2g\abs{QB}}
	 \begin{pmatrix} -S_{-s} \hat{a} + S_{s} \hat{a}^\dagger & 0 \\ 
	 0  & S_{-s} \hat{a} - S_{s} \hat{a}^\dagger\end{pmatrix}
	+ \Pi_z  \begin{pmatrix}  \sigma_3 & 0 \\ 0 & -\sigma_3 \end{pmatrix}
	- m \begin{pmatrix} 0 & e^{2i\theta_m} \\ e^{-2i \theta_m} & 0 \end{pmatrix}
	+ \dot{\theta}_5 \begin{pmatrix} -1 & 0 \\ 0 & 1 \end{pmatrix}
	\right] \psi,
\end{align}
where
\begin{align}
	\bar{x}_s = \sqrt{g\abs{QB}} \left(x - s\frac{p_y}{g\abs{QB}}\right),
	\quad
	S_\pm = \frac{1}{2}\left(\sigma_1 \pm i \sigma_2\right),
	\quad
	\hat{a} = \frac{1}{\sqrt{2}}\left(\partial_{\bar{x}_s} + \bar{x}_s\right),
	\quad
	\hat{a}^\dagger = \frac{1}{\sqrt{2}}\left(-\partial_{\bar{x}_s} + \bar{x}_s\right),
\end{align}
and $s = \mathrm{sgn}(QB)$. We expand the modes as
\begin{align}
	\psi = \sum_{n, s', \lambda} \psi_{n, s'}^{\left(\lambda\right)} h_{n}(\bar{x}_{s})\chi_{s'}^{\left(\lambda\right)},
	\label{eq:fermion_decomposition}
\end{align}
where $h_n$ is related to the Hermite polynomial $H_n$ as
\begin{align}
	h_n(\bar{x}_s) &= \left(\frac{g\abs{QB}}{\pi}\right)^{1/4}\left(\frac{1}{2^n n!}\right)^{1/2} 
	e^{-\bar{x}_s^2/2}H_n(\bar{x}_s),
\end{align}
 and satisfies
\begin{align}
	\hat{a} h_n = \sqrt{n}\, h_{n-1},
	\quad
	\hat{a}^\dagger h_n = \sqrt{n+1}\,h_{n+1},
	\quad
	\int \dd x h_n(\bar{x}_s) h_{n'}(\bar{x}_s) = \delta_{n n'}.
\end{align}
The spinor $\chi$ is given by
\begin{align}
	\chi_+^{\left({L}\right)} = (1~~0~~0~~0)^{T},
	\quad
	\chi_-^{\left({L}\right)} = (0~~1~~0~~0)^{T},
	\quad
	\chi_+^{\left({R}\right)} = (0~~0~~1~~0)^{T},
	\quad
	\chi_-^{\left({R}\right)} = (0~~0~~0~~1)^{T}.
\end{align}
The operator $S_{s} \hat{a}^\dagger$ changes $s'=-s \to s$ and $n \to n+1$,
while $S_{-s} \hat{a}$ changes $s'=s \to -s$ and $n \to n-1$,
and the combination $2n+1 - ss'$ is invariant under both operations.
It then follows that only the modes with the same value of $2n + 1 - s s' $ mix with each other.
The mode with $2n + 1 - s s' = 0$, \textit{i.e.}, $n = 0$ and $s' = s$ corresponds to the lowest Landau level,
while the others are the higher Landau levels.

%%%%%
\paragraph{Lowest Landau level.}
%%%%%
The equation of motion of the lowest Landau level is given by
\begin{align}
	0 &= \begin{pmatrix} i\partial_0 + s\Pi_z - \dot{\theta}_5 & - me^{2i\theta_m} \\
	-m e^{-2i\theta_m} & i\partial_0 - s\Pi_z + \dot{\theta}_5 \end{pmatrix}
	\begin{pmatrix}
	\psi^{(L)}_{0,s} \\
	\psi^{(R)}_{0,s}
	\end{pmatrix}.
\end{align}
In the Hamiltonian basis, we eliminate $c_5$ by a chiral rotation.
The solutions of the Dirac equation for constant $\phi$ and $A_z$ are then given by
\begin{align}
	u_0 = \frac{e^{-i\gamma_5\theta_{5+m}}}
	{\sqrt{2\Omega_0 \left(\Omega_0 + s {\Pi}_z\right)}}
	\left[m \chi_s^{(L)} + \left(\Omega_0 + s\Pi_z\right) \chi_s^{(R)}\right]h_0,
	\quad
	v_0 = \frac{e^{-i\gamma_5\theta_{5+m}}}
	{\sqrt{2\Omega_0 \left(\Omega_0 + s {\Pi}_z\right)}}
	\left[\left(\Omega_0 + s\Pi_z\right)\chi_s^{(L)} -m \chi_s^{(R)}\right]h_0,
\end{align}
where $\Omega_0 = \sqrt{\Pi_z^2 + m^2}$.
After turning on the background fields, the Bogoliubov coefficients evolve as
\begin{align}
	\dot{\alpha}_0 &= i\dot{\theta}_{5+m}  \frac{s {\Pi}_z}{\Omega_0} \alpha_0 
	- \left(s\frac{m \dot{{\Pi}}_z}{2\Omega_0^2} + i \dot{\theta}_{5+m}  \frac{m}{\Omega_0}\right) e^{2i\Theta_0} \beta_0, \\
	\dot{\beta}_0 &= -i\dot{\theta}_{5+m}  \frac{s {\Pi}_z}{\Omega_0} \beta_0 
	+ \left(s\frac{m \dot{{\Pi}}_z}{2\Omega_0^2} - i \dot{\theta}_{5+m}  \frac{m}{\Omega_0}\right) e^{-2i\Theta_0} \alpha_0,
\end{align}
where $\Theta_0 = \int^t \dd t\,\Omega_0$.
We solve this equation with $\alpha_0 = 1$ and $\beta_0 =0$ as the initial condition.
The lowest Landau level is then quantized in the Hamiltonian basis as
\begin{align}
	\psi_0 &= \sum_{\lambda} \psi_{0, s}^{\left(\lambda\right)} h_{0}\chi_{s}^{\left(\lambda\right)}
	=
	e^{i\gamma_5 \theta_5}
	\left[\left(\alpha_0 b_0 - \beta_0^* d_0^\dagger\right)u_0 e^{-i\Theta_0}
	+  \left(\beta_0 b_0 + \alpha_0^* d_0^\dagger\right)v_0 e^{i\Theta_0}
	\right].
\end{align}

In the \FCBnsp, the solution of the Dirac equation for constant $\dot{\phi}$ and $A_z$ 
is given by
\begin{align}
	\tilde{u}_0 &= \frac{m \chi_s^{(L)} +\left( \Omega_{0,s} + s \Pi_z - \dot{\theta}_{5+m}\right) \chi_s^{(R)}}
	{\sqrt{2\Omega_{0,s} \left(\Omega_{0,s} + s \Pi_z- \dot{\theta}_{5+m}\right)}}h_0,
	\quad
	\tilde{v}_0 = 
	\frac{\left(\Omega_0 + s {\Pi}_z - \dot{\theta}_{5+m}\right)\chi_s^{(L)} - m \chi_s^{(R)}}
	{\sqrt{2\Omega_{0,s} \left(\Omega_{0,s} + s \Pi_z- \dot{\theta}_{5+m}\right)}}h_0,
\end{align}
where $\Omega_{0,s} = \sqrt{(\Pi_z - s\dot{\theta}_{5+m})^2 + m^2}$.
The Bogoliubov coefficients evolve as
\begin{align}
	\dot{\tilde\alpha}_0 &= 
	- \frac{m \left(s\dot{{\Pi}}_z - \ddot{\theta}_{5+m}\right)}{2\Omega_{0,s}^2} e^{2i\Theta_{0,s}} \tilde\beta_0, 
	\quad
	\dot{\tilde\beta}_0 =
	+ \frac{m \left(s\dot{{\Pi}}_z - \ddot{\theta}_{5+m}\right)}{2\Omega_{0,s}^2} e^{-2i\Theta_{0,s}} \tilde\alpha_0,
\end{align}
where $\Theta_{0,s} = \int^t \dd t \, \Omega_{0,s}$.
The fermion is then quantized as
\begin{align}
	\psi_0 &= 
	e^{-i\gamma_5 \theta_m}
	\left[\left(\tilde\alpha_0 {b}_0 - \tilde{\beta}_0^* d_0^\dagger\right)\tilde{u}_0 e^{-i\Theta_{0,s}}
	+  \left(\tilde{\beta}_0 {b}_0 + \tilde{\alpha}_0^* {d}_0^\dagger\right)\tilde{v}_0 e^{i\Theta_{0,s}}
	\right].
\end{align}
The creation and annihilation operators satisfy
\begin{align}
	\left\{b_{0}(\vec{p}), b_{0}^\dagger(\vec{p}')\right\}
	&= \left\{d_{0}(\vec{p}), d_{0}^\dagger(\vec{p}')\right\}
	=(2\pi)^2\delta^{(2)}(\vec{p}-\vec{p}'),
	\\
	\left\{b_{0}(\vec{p}), d_{0}^\dagger(\vec{p}')\right\}
	&= \left\{b_{0}(\vec{p}), d_{0}(\vec{p}')\right\}
	= 0,
\end{align}
and $B_0$, $D_0$, $\tilde{B}_0$ and $\tilde{D}_0$ satisfy the same equal-time anti-commutators,
which one can show by noting that
\begin{align}
	\abs{\alpha_0}^2 + \abs{\beta_0}^2
	= \abs{\tilde{\alpha}_0}^2 + \abs{\tilde{\beta}_0}^2 = 1.
\end{align}
%%

%%%%%
\paragraph{Higher Landau levels.}
%%%%%

The equation of motion of the higher Landau levels is given by
\begin{align}
	0 = \begin{pmatrix}
	i \partial_0 + s\Pi_z - \dot{\theta}_5 & i m_B & -m e^{2i\theta_m} & 0 \\
	-i m_B & i \partial_0 - s\Pi_z - \dot{\theta}_5 & 0 & -m e^{2i\theta_m} \\
	- m e^{-2i\theta_m} & 0 & i\partial_0 - s\Pi_z + \dot{\theta}_5 & -i m_B \\
	0 & -m e^{-2i\theta_m} & i m_B & i\partial_0 + s\Pi_z + \dot{\theta}_5
	\end{pmatrix}
	\begin{pmatrix}
	\psi_{n,s}^{(L)} \\ \psi_{n-1,-s}^{(L)} \\ \psi_{n,s}^{(R)} \\ \psi_{n-1,-s}^{(R)}
	\end{pmatrix},
\end{align}
for $n = 1,2,...$ and $m_B = \sqrt{2ng\abs{QB}}$.
In the Hamiltonian basis, the solutions of the Dirac equation for constant $\phi$ and $A_z$ are given by
\begin{align}
	u_{n,1} &= \frac{e^{-i\gamma_5 \theta_{5+m}}}{N}
	\left[
	im_B\left(- m_T \chi_s^{(L)} + \left(\Omega_n + s \Pi_z\right)\chi_{s}^{(R)}  \right)h_n
	+\left(m + m_T\right)\left(\left(\Omega_n + s \Pi_z\right)\chi_{-s}^{(L)} + m_T \chi_{-s}^{(R)}
	\right) h_{n-1}
	\right],
	\\
	u_{n,2} &= \frac{ie^{-i\gamma_5 \theta_{5+m}}}{N}
	\left[\left(m + m_T\right)\left(m_T\chi_s^{(L)} + \left(\Omega_n + s \Pi_z\right) \chi_{s}^{(R)} \right)h_n
	+im_B\left( \left(\Omega_n + s \Pi_z\right)\chi_{-s}^{(L)}
	- m_T \chi_{-s}^{(R)}\right)h_{n-1}\right],
	\\
	v_{n,1} &= \frac{i e^{-i\gamma_5 \theta_{5+m}}}{N}
	\left[-im_B\left( \left(\Omega_n + s \Pi_z\right) \chi_s^{(L)} + m_T \chi_{s}^{(R)} \right)h_n
	+ \left(m + m_T\right)\left(-m_T \chi_{-s}^{(L)}+\left(\Omega_n + s \Pi_z\right)\chi_{-s}^{(R)}\right)h_{n-1}\right],
	 \\
	v_{n,2} &= \frac{i e^{-i\gamma_5 \theta_{5+m}}}{N}
	\left[\left(m + m_T\right) \left( \left(\Omega_n + s \Pi_z\right) \chi_s^{(L)} -m_T \chi_{s}^{(R)} \right)h_n
	 -i m_B
	 \left(m_T \chi_{-s}^{(L)} + \left(\Omega_n + s \Pi_z\right)\chi_{-s}^{(R)}\right)h_{n-1}\right],
\end{align}
where $m_T^2 = m^2 + m_B^2$ and $\Omega_n = \sqrt{\Pi_z^2 + m_T^2}$.
The normalization factor is given by
\begin{align}
	N = 2\sqrt{\Omega_n \left(\Omega_n + s \Pi_z\right) m_T\left(m + m_T\right)}.
\end{align}
After turning on the time dependence of the background fields,
the time evolution of the Bogoliubov coefficients is given by
\begin{align}
	\begin{pmatrix}
	\dot{\alpha}_{n,1} \\
	\dot{\alpha}_{n,2} \\
	\dot{\beta}_{n,1} \\
	\dot{\beta}_{n,2} 
	\end{pmatrix}
	=&
	\left[
	i\dot{\theta}_{5+m}
	\begin{pmatrix}
	- \frac{m}{m_T}\frac{s\Pi_z}{\Omega_n} & \frac{m_B}{m_T} & \frac{m}{\Omega_n}e^{2i\Theta_n} & 0 \\
	\frac{m_B}{m_T} & \frac{m}{m_T}\frac{s\Pi_z}{\Omega_n} & 0 & -\frac{m}{\Omega_n}e^{2i\Theta_n} \\
	\frac{m}{\Omega_n}e^{-2i\Theta_n} & 0 & \frac{m}{m_T}\frac{s\Pi_z}{\Omega_n} & \frac{m_B}{m_T} \\
	0 & -\frac{m}{\Omega_n}e^{-2i\Theta_n} & \frac{m_B}{m_T} & -\frac{m}{m_T}\frac{s\Pi_z}{\Omega_n}
	\end{pmatrix} % \right. \nonumber \\
	%\left.
	  + \frac{ s m_T \dot{\Pi}_z}{2\Omega_n^2}
	\begin{pmatrix}
	0 & 0 & -e^{2i\Theta_n} & 0 \\
	0 & 0 & 0 & -e^{2i\Theta_n} \\
	e^{-2i\Theta_n} & 0 & 0 & 0 \\
	0 & e^{-2i\Theta_n} & 0 & 0
	\end{pmatrix}
	\right]
	\begin{pmatrix}
	{\alpha}_{n,1} \\
	{\alpha}_{n,2} \\
	{\beta}_{n,1} \\
	{\beta}_{n,2} 
	\end{pmatrix},
\end{align}
where $\Theta_n = \int^t dt\,\Omega_n$.
Notice that this is equivalent to the case without the magnetic field after replacing $m_B \to p_T$.
Finally the fermion is quantized as
\begin{align}
	\psi_n &= 
	\sum_{\lambda} \left[\psi_{n, s}^{\left(\lambda\right)} h_{n}\chi_{s}^{\left(\lambda\right)} 
	+ \psi_{n-1, -s}^{\left(\lambda\right)} h_{n-1}\chi_{-s}^{\left(\lambda\right)}\right]
	\nonumber \\
	&= 
	\sum_{\lambda, \lambda' = 1,2} \left[ u_{n,\lambda} e^{-i\Theta_n}
	\left(\alpha_{n,\lambda}^{(\lambda')} b_{n,\lambda'} - 
	\left(-1\right)^{\lambda+\lambda'}{\beta_{n,\lambda}^{(\lambda')}}^* d_{n,\lambda'}^\dagger \right)
	+
	v_{n,\lambda} e^{i\Theta_n}
	\left(\beta_{n,\lambda}^{(\lambda')} b_{n,\lambda'} 
	+ \left(-1\right)^{\lambda+\lambda'}{\alpha_{n,\lambda}^{(\lambda')}}^* d_{n,\lambda'}^\dagger \right)
	\right].
\end{align}
Here the superscript indicates the initial conditions for $\alpha$ and $\beta$, that is,
\begin{align}
	\alpha_{n,1}^{(1)} &= 1,
	\quad
	\alpha_{n,2}^{(1)} = \beta_{n,1}^{(1)} = \beta_{n,2}^{(1)} = 0, \\
	\alpha_{n,2}^{(2)} &= 1,
	\quad
	\alpha_{n,1}^{(2)} = \beta_{n,1}^{(2)} = \beta_{n,2}^{(2)} = 0, 
\end{align}
at the initial time.

In the \FCBnsp, the solutions of the Dirac equation for constant $\dot{\phi}$ and $A_z$
are given by
\begin{align}
	\tilde{u}_{n,+} &= \frac{1}{N_+}
	\left[-im_B\left( \left(\Omega_{n,+} + \Pi_B + \dot{\theta}_{5+m}\right) \chi_s^{(L)} + m \chi_s^{(R)}\right)h_n
	+ \left(\Pi_B + s\Pi_z\right)\left(
	\left(\Omega_{n,+} + \Pi_B + \dot{\theta}_{5+m}\right) \chi_{-s}^{(L)} + m \chi_{-s}^{(R)}\right)h_{n-1}
	\right], \\
	\tilde{u}_{n,-} &= \frac{1}{N_-}
	\left[
	im_B \left(m \chi_s^{(L)}  + \left(\Omega_{n,-} + \Pi_B - \dot{\theta}_{5+m}\right)\chi_s^{(R)}\right)h_n
	+\left(\Pi_B - s\Pi_z \right) 
	\left(m\chi_{-s}^{(L)} + \left(\Omega_{n,-} + \Pi_B - \dot{\theta}_{5+m}\right)\chi_{-s}^{(R)}\right)h_{n-1}
	\right]
	\\
	\tilde{v}_{n,+} &= \frac{1}{N_+}
	\left[im_B\left(m \chi_{s}^{(L)} - \left(\Omega_{n,+} + \Pi_B + \dot{\theta}_{5+m}\right)\chi_s^{(R)}\right) h_n
	+\left(\Pi_B + s\Pi_z \right)\left(-m \chi_{-s}^{(L)} + \left(\Omega_{n,+} + \Pi_B + \dot{\theta}_{5+m}\right)
	\chi_{-s}^{(R)}\right)h_{n-1}
	\right],
	\\
	\tilde{v}_{n,-} &= \frac{1}{N_-}
	\left[
	im_B \left(-\left(\Omega_{n,-} + \Pi_B - \dot{\theta}_{5+m}\right)\chi_s^{(L)} + m\chi_s^{(R)}\right)h_n
	+\left(\Pi_B - s\Pi_z \right) \left(-\left(\Omega_{n,-} + \Pi_B - \dot{\theta}_{5+m}\right)\chi_{-s}^{(L)}
	+ m \chi_{-s}^{(R)}\right)
	\right],
\end{align}
where $\Pi_B = \sqrt{\Pi_z^2 + m_B^2}$, $\Omega_{n,\pm} = \sqrt{(\Pi_B \pm \dot{\theta}_{5+m})^2+m^2}$,
and $N_\pm = 2 \sqrt{\Pi_B \left(\Pi_B \pm s\Pi_z\right) \Omega_{n,\pm} \left(\Omega_{n,\pm} + \Pi_B  
\pm \dot{\theta}_{5+m}\right)}$.
The time evolution equations for the Bogoliubov coefficients are given by
\begin{align}
	\begin{pmatrix}
	\dot{\tilde\alpha}_{n,+} \\
	\dot{\tilde\alpha}_{n,-} \\
	\dot{\tilde\beta}_{n,+} \\
	\dot{\tilde\beta}_{n,-}
	\end{pmatrix}
	&= \begin{pmatrix}
	0 & A_ne^{i(\Theta_{n,+} - \Theta_{n,-})} & -B_{n,+} e^{2i\Theta_{n,+}} & -C_n e^{i(\Theta_{n,+} + \Theta_{n,-})} \\
	-A_n e^{-i(\Theta_{n,+} - \Theta_{n,-})} & 0 & -C_ne^{i(\Theta_{n,+} + \Theta_{n,-})} & -B_{n,-}e^{2i\Theta_{n,-}} \\
	B_{n,+} e^{-2i\Theta_{n,+}} & C_n e^{-i(\Theta_{n,+} + \Theta_{n,-})}& 0 & A_n e^{-i(\Theta_{n,+} - \Theta_{n,-})}  \\
	C_n e^{-i(\Theta_{n,+} + \Theta_{n,-})} & B_{n,-} e^{-2i\Theta_{n,-}} & -A_n e^{i(\Theta_{n,+} - \Theta_{n,-})}  & 0
	\end{pmatrix}
	\begin{pmatrix}
	\tilde{\alpha}_{n,+} \\
	\tilde{\alpha}_{n,-} \\
	\tilde{\beta}_{n,+} \\
	\tilde{\beta}_{n,-}
	\end{pmatrix},
	\label{eq:eom_alpha_beta}
\end{align}
where
\begin{align}
	A_n &= s\frac{gQE m_B m}{4\Pi_B^2 \sqrt{\Omega_{n,+} \Omega_{n,-}}}
	\left[\frac{\Omega_{n,+} + \Omega_{n,-} + 2\Pi_B}{\sqrt{\left(\Omega_{n,+} + \Pi_B + \dot{\theta}_{5+m}\right)
	\left(\Omega_{n,-} + \Pi_B - \dot{\theta}_{5+m}\right)}}\right], \\
	B_{n,\pm} &= \pm \frac{m}{2\Omega_{n,\pm}^2}\left(\frac{\Pi_z}{\Pi_B} gQE \pm \ddot{\theta}_{5+m}\right), 
	\quad
	C_n= s\frac{gQEm_B \left[ \left(\Omega_{n,+} + \Pi_B + \dot{\theta}_{5+m}\right)
	\left(\Omega_{n,-} + \Pi_B - \dot{\theta}_{5+m}\right) - m^2\right]}
	{4\Pi_B^2 \sqrt{\Omega_{n,+}\Omega_{n,-} 
	\left(\Omega_{n,+} + \Pi_B + \dot{\theta}_{5+m}\right)\left(\Omega_{n,-} + \Pi_B - \dot{\theta}_{5+m}\right)}}.
\end{align}
We solve these equations with two different initial conditions,
\begin{align}
	\tilde\alpha_{n,+}^{(+)} &= 1,
	\quad
	\tilde\alpha_{n,-}^{(+)} = \tilde\beta_{n,+}^{(+)} = \tilde\beta_{n,-}^{(+)} = 0,
	\qquad\mathrm{and}\qquad
	\tilde\alpha_{n,-}^{(-)} = 1,
	\quad
	\tilde\alpha_{n,+}^{(-)} =\tilde \beta_{n,+}^{(-)} = \tilde\beta_{n,-}^{(-)} = 0.
\end{align}
The Dirac field is quantized as
\begin{align}
	\psi_n =  
	e^{-i\theta_m\gamma_5}\sum_{\lambda \lambda' = \pm}
	\left[\tilde{u}_{n,\lambda} e^{-i\Theta_{n,\lambda}}
	\left(\tilde{\alpha}_{n,\lambda}^{(\lambda')}{b}_{n,\lambda'}
	-\tilde{\beta}_{n,\lambda}^{(\lambda')*}{d}_{n,\lambda'}^{\,\dagger}\right)
	+ \tilde{v}_{n,\lambda} e^{i\Theta_{n,\lambda}}\left(\tilde{\beta}_{n,\lambda}^{(\lambda')}{b}_{n,\lambda'} 
	+ \tilde{\alpha}_{n,\lambda}^{(\lambda')*} {d}_{n,\lambda'}^{\,\dagger}\right)
	\right].
\end{align}
Finally, the creation and annihilation operators satisfy
\begin{align}
	\left\{b_{n_1,\lambda_1}(\vec{p}), b_{n_2,\lambda_2}^\dagger(\vec{p}')\right\}
	&= \left\{d_{n_1, \lambda_1}(\vec{p}), d_{n_2, \lambda_2}^\dagger(\vec{p}')\right\}
	=(2\pi)^2\delta_{\lambda_1 \lambda_2}\delta_{n_1 n_2}\delta^{(2)}(\vec{p}-\vec{p}'),
	\\
	\left\{b_{n_1, \lambda_1}(\vec{p}), d_{n_2, \lambda_2}^\dagger(\vec{p}')\right\}
	&= \left\{b_{n_1, \lambda_1}(\vec{p}), d_{n_2, \lambda_2}(\vec{p}')\right\}
	= 0,
\end{align}
and $B$, $D$, $\tilde{B}$ and $\tilde{D}$ satisfy the same equal-time anti-commutators,
which one can show in the same as the case without the magnetic field.

%%%%%
\paragraph{Summary.}
%%%%%
In summary, in the presence of the magnetic field, the fermion is quantized as
\begin{align}
	\psi =& \int \frac{\dd p_y \dd p_z}{(2\pi)^2}e^{i(p_y y + p_z z)}
	e^{i\gamma_5 \theta_5}
	\left[\left(\alpha_0 b_0 - \beta_0^* d_0^\dagger\right)u_0 e^{-i\Theta_0}
	+ \left(\beta_0 b_0 + \alpha_0^* d_0^\dagger\right)v_0 e^{i\Theta_0}
	\right. \nonumber \\
	&\left. + \sum_{n, \lambda, \lambda'}
	\left[ u_{n,\lambda} e^{-i\Theta_n}
	\left(\alpha_{n,\lambda}^{(\lambda')} b_{n,\lambda'} - 
	\left(-1\right)^{\lambda+\lambda'}{\beta_{n,\lambda}^{(\lambda')}}^* d_{n,\lambda'}^\dagger \right)
	+
	v_{n,\lambda} e^{i\Theta_n}
	\left(\beta_{n,\lambda}^{(\lambda')} b_{n,\lambda'} 
	+ \left(-1\right)^{\lambda+\lambda'}{\alpha_{n,\lambda}^{(\lambda')}}^* d_{n,\lambda'}^\dagger \right)
	\right]
	\right],
\end{align}
in the Hamiltonian basis, and
\begin{align}
	\psi =& \int \frac{\dd p_y \dd p_z}{(2\pi)^2}e^{i(p_y y + p_z z)}
	e^{-i\gamma_5 \theta_m}
	\left[\left(\tilde{\alpha}_0 {b}_0 -\tilde{\beta}_0^* {d}_0^\dagger\right)\tilde{u}_0 e^{-i\Theta_{0,s}}
	+ \left(\tilde\beta_0 {b}_0 + \tilde\alpha_0^* {d}_0^\dagger\right)\tilde{v}_0 e^{i\Theta_{0,s}}
	\right. \nonumber \\
	&\left. + \sum_{n, \lambda, \lambda'}\left[\tilde{u}_{n,\lambda} e^{-i\Theta_{n,\lambda}}
	\left(\tilde{\alpha}_{n,\lambda}^{(\lambda')}{b}_{n,\lambda'}
	-\tilde{\beta}_{n,\lambda}^{(\lambda')*}{d}_{n,\lambda'}^{\,\dagger}\right)
	+ \tilde{v}_{n,\lambda} e^{i\Theta_{n,\lambda}}\left(\tilde{\beta}_{n,\lambda}^{(\lambda')}{b}_{n,\lambda'} 
	+ \tilde{\alpha}_{n,\lambda}^{(\lambda')*} {d}_{n,\lambda'}^{\,\dagger}\right)
	\right]
	\right],
\end{align}
in the \FCBnsp, respectively.

%%%%%
\subsection{Physical quantities}
%%%%%
For convenience, we summarize the physical quantities expressed in terms of the Bogoliubov coefficients in
both the Hamiltonian and the {\FCBsnsp} in this appendix.
We focus on the energy density, the induced current, the axial charge and the chiral mass operator,
which are defined as
\begin{align}
	\rho_\psi &\equiv \frac{1}{2\mathrm{Vol}(\mathbb{R}_3)}\int \dd^3x 
	\left\langle \left[\psi^\dagger, \left(i\partial_0 + \dot{\theta}_5 \gamma_5\right) \psi \right] \right\rangle,
	\quad
	\left\langle J_z \right\rangle 
	\equiv \frac{1}{2\mathrm{Vol}(\mathbb{R}_3)}\int \dd^3x \left\langle \left[\bar{\psi}, \gamma^3 \psi \right] \right\rangle, \\
	q_5 &\equiv \frac{1}{2\mathrm{Vol}\left(\mathbb{R}_3\right)}\int \dd^3x
	\left\langle 
	\left[ {\psi}^\dagger, \gamma_5 \psi\right]
	\right\rangle,
	\quad
	\left\langle \bar{\psi} e^{2i\theta_m \gamma_5} i\gamma_5 \psi \right\rangle
	\equiv \frac{1}{2\mathrm{Vol}\left(\mathbb{R}_3\right)}\int \dd^3x
	\left\langle 
	\left[ \bar{\psi}, i\gamma_5 e^{2i\theta_m \gamma_5} \psi\right]
	\right\rangle.
\end{align}
Below give the expressions both cases, without and with the magnetic field, separately.

%%%%%
\paragraph{Without magnetic field.}
%%%%%
The averaged energy density of the fermion is given as the expectation value of the Hamiltonian as
$\rho_\psi \equiv {\left\langle H_\psi \right\rangle}/{\mathrm{Vol}(\mathbb{R}_3)}$.
This is given by
\begin{align}
	\rho_\psi
	&= 
	\int \frac{\dd^3 p}{(2\pi)^3}\Omega \sum_{\lambda, \lambda'}
	\left[\abs{\beta_{\lambda}^{(\lambda')}}^2 - \abs{\alpha_{\lambda}^{(\lambda')}}^2\right]
	\nonumber \\
	&= \int \frac{\dd^3 p}{(2\pi)^3}
	\sum_{\lambda,\lambda'}\left[\frac{\Pi^2 + m^2 +\lambda \dot{\theta}_{5+m}\Pi}{\Omega_{\lambda}}
	\left(\abs{\tilde{\beta}_{\lambda}^{(\lambda')}}^2 - \abs{\tilde{\alpha}_{\lambda}^{(\lambda')}}^2\right)
	+ \frac{2m\dot{\theta}_{5+m}}{\Omega_{\lambda}}
	\mathrm{Re}\left[\tilde{\alpha_{\lambda}^{(\lambda')}}^* \tilde{\beta}_{\lambda}^{(\lambda')} 
	e^{2i\Theta_{\lambda}}\right]
	\right],
\end{align}
where here and in the following, the first line is computed in the Hamiltonian basis and the second line is computed in the \FCBnsp.
The induced current is given by
\begin{align}
	\langle J_z \rangle
	=
	\int \frac{\dd^3 p}{(2\pi)^3}
	&\sum_{\lambda,\lambda'}
	\left[
	\frac{\Pi_z}{\Omega}
	\left(\abs{\beta_{\lambda}^{(\lambda')}}^2 - \abs{\alpha_{\lambda}^{(\lambda')}}^2\right)
	+ \frac{2m_T}{\Omega}\mathrm{Re}\left[\alpha_{\lambda}^{(\lambda')*} \beta_{\lambda}^{(\lambda')}e^{2i\Theta}\right]
	\right]
	\nonumber \\
	=
	\int \frac{\dd^3  p}{(2\pi)^3}
	&\sum_{\lambda'}
	\left\{\sum_\lambda\left[
	\frac{\lambda \Pi_z}{\Pi}\frac{\Pi+\lambda \dot{\theta}_{5+m}}{\Omega_\lambda}
	\left(\abs{\tilde\beta_{\lambda}^{(\lambda')}}^2 - \abs{\tilde\alpha_{\lambda}^{(\lambda')}}^2\right)
	+ \frac{2\Pi_z}{\Pi}\frac{m}{\Omega_\lambda}
	\mathrm{Re}\left[\tilde{\alpha}_{\lambda}^{(\lambda')*}\tilde{\beta}_{\lambda}^{(\lambda')}\right]
	\right]
	\right. \nonumber \\ &~~~~~\left.
	+\frac{p_T}{\Pi}\frac{m(\Omega_+ - \Omega_- + 2\dot{\theta}_{5+m})}
	{\tilde{N}}
	\mathrm{Re}\left[e^{i(\Theta_+ - \Theta_-)}
	\left(\tilde{\beta}_-^{(\lambda')*} \tilde{\beta}_+^{(\lambda')}
	- \tilde{\alpha}_+^{(\lambda')*} \tilde{\alpha}_-^{(\lambda')}\right)\right]
	\right. \nonumber \\ &~~~~~\left.
	+\frac{2p_T}{\Pi}\frac{(\Omega_+ + \Pi + \dot{\theta}_{5+m})(\Omega_- + \Pi - \dot{\theta}_{5+m}) + m^2}
	{\tilde{N}}
	\mathrm{Re}\left[e^{i(\Theta_+ + \Theta_-)}\left(\tilde{\alpha}_+^{(\lambda')*}\tilde{\beta}_-^{(\lambda')}
	+ \tilde{\alpha}_-^{(\lambda')*}\tilde{\beta}_+^{(\lambda')}\right)\right]
	\right\},
\end{align}
where
\begin{align}
	\tilde{N} = \sqrt{\Omega_+\Omega_-(\Omega_+ + \Pi + \dot{\theta}_{5+m})(\Omega_- + \Pi - \dot{\theta}_{5+m})}.
\end{align}
The axial charge is given by
\begin{align}
	q_5
	&=
	\int\frac{\dd^3  p}{(2\pi)^3}
	\sum_{\lambda}\left[
	\frac{m}{m_T}\frac{\Pi_z}{\Omega}
	\left(
	\abs{\alpha_{1}^{(\lambda')}}^2 - \abs{\beta_{1}^{(\lambda')}}^2
	- \abs{\alpha_{2}^{(\lambda)}}^2 + \abs{\beta_{2}^{(\lambda)}}^2
	\right)
	\right. \nonumber \\ &~~~~~~~~~~~~~~~~~~~~~~~~~~~~~~~~~~~~~~~~\left.
	- \frac{2p_T}{m_T}\mathrm{Re}\left[\alpha_{1}^{(\lambda)*}\alpha_{2}^{(\lambda)}
	+\beta_{1}^{(\lambda)*}\beta_{2}^{(\lambda)}\right]
	-\frac{2m}{\Omega}\mathrm{Re}\left[\left(\alpha_{1}^{(\lambda)*}\beta_{1}^{(\lambda)}
	- \alpha_{2}^{(\lambda)*}\beta_{2}^{(\lambda)}\right)e^{2i\Theta}
	\right]
	\right]
	\nonumber \\
	&= 
	\int\frac{\dd^3  p}{(2\pi)^3}
	\sum_{\lambda,\lambda'}
	\left[
	\frac{\lambda \Pi + \dot{\theta}_{5+m}}{\Omega_{\lambda}}
	\left(\abs{\tilde{\alpha}_{\lambda}^{(\lambda')}}^2 - \abs{\tilde{\beta}_{\lambda}^{(\lambda')}}^2\right)
	-\frac{2m}{\Omega_{\lambda}}
	\mathrm{Re}\left[
	\tilde{\alpha}_{\lambda}^{(\lambda')}\tilde{\beta}_{\lambda}^{(\lambda')}e^{2i\Theta_{\lambda}}
	\right]
	\right].
\end{align}
Finally the chiral mass operator is given by
\begin{align}
	\left\langle \bar{\psi} e^{2i\theta_m \gamma_5} i\gamma_5 \psi \right\rangle
	 &= 2\int\frac{\dd^3 p}{(2\pi)^3}\sum_{\lambda}\mathrm{Im}\left[
	 \left(\alpha_{1}^{(\lambda)*} \beta_{1}^{(\lambda)} 
	 - \alpha_{2}^{(\lambda)*}\beta_{2}^{(\lambda)}\right)e^{2i\Theta}
	 \right]
	 \nonumber \\
	 &=2\int\frac{\dd^3 p}{(2\pi)^3}
	\sum_{\lambda,\lambda'}\mathrm{Im}\left[
	 \tilde{\alpha}_{\lambda}^{(\lambda')*}\tilde{\beta}_{\lambda}^{(\lambda')} e^{2i\Theta_{\lambda}}
	 \right].
\end{align}
Notice that we do not perform the regularization and renormalization of these quantities in this appendix.

%%%%%
\paragraph{With magnetic field.}
%%%%%
The averaged energy density of the fermion is given by\footnote{
	We note that $\int_{-\infty}^{\infty} \dd x \dd p_y \abs{h_n(\bar{x}_s)}^2 = 
	\int_{-\infty}^{\infty} \dd p_y = g\abs{QB} \int_{-\infty}^{\infty} \dd x$.
}
\begin{align}
	\rho_\psi
	= 
	\frac{g\abs{QB}}{4\pi^2}\int \dd p_z &\left[\Omega_0 \left(\abs{\beta_0}^2-\abs{\alpha_0}^2\right)
	+ \sum_{n,\lambda,\lambda'} \Omega_n 
	\left(\abs{\beta_{n,\lambda}^{(\lambda')}}^2 - \abs{\alpha_{n,\lambda}^{(\lambda')}}^2\right)\right]
	\nonumber \\
	= 
	\frac{g\abs{QB}}{4\pi^2}\int \dd p_z&\left\{
	\frac{\Pi_z^2 + m^2 - s \dot{\theta}_{5+m}\Pi_z}{\Omega_{0,s}}
	\left(\abs{\tilde{\beta}_0}^2 - \abs{\tilde{\alpha}_0}^2\right)
	+ \frac{2m\dot{\theta}_{5+m}}{\Omega_{0,s}}\mathrm{Re}\left[\tilde{\alpha_0}^* \tilde{\beta}_0 e^{2i\Theta_{0,s}}\right]
	\right. \nonumber \\ &\left.
	+\sum_{n,\lambda,\lambda'}\left[\frac{\Pi_B^2 + m^2 +\lambda \dot{\theta}_{5+m}\Pi_B}{\Omega_{n,\lambda}}
	\left(\abs{\tilde{\beta}_{n,\lambda}^{(\lambda')}}^2 - \abs{\tilde{\alpha}_{n,\lambda}^{(\lambda')}}^2\right)
	+ \frac{2m\dot{\theta}_{5+m}}{\Omega_{n,\lambda}}
	\mathrm{Re}\left[\tilde{\alpha_{n,\lambda}^{(\lambda')}}^* \tilde{\beta}_{n,\lambda}^{(\lambda')} 
	e^{2i\Theta_{n,\lambda}}\right]
	\right]
	\right\}.
\end{align}
The induced current is given by
\begin{align}
	\langle J_z \rangle 
	= \frac{gQB}{4\pi^2} \int \dd p_z
	&\left\{\frac{s\Pi_z}{\Omega_0}\left[\abs{\beta_0}^2-\abs{\alpha_0}^2\right]
	+ \frac{2m}{\Omega_0}\mathrm{Re}\left[\alpha_0^* \beta_0 e^{2i\Theta_0}\right]
	\right. \nonumber \\ &\left.
	+\sum_{n,\lambda,\lambda'}\left[
	\frac{s\Pi_z}{\Omega_n}\left[\abs{\beta_{n,\lambda}^{(\lambda')}}^2-\abs{\alpha_{n,\lambda}^{(\lambda')}}^2\right]
	+ \frac{2m_T}{\Omega_n}
	\mathrm{Re}\left[\alpha_{n,\lambda}^{(\lambda')*} \beta_{n,\lambda}^{(\lambda')} e^{2i\Theta_0}\right]
	\right]
	\right\}
	\nonumber \\
	= \frac{gQB}{4\pi^2}\int \dd p_z
	&\left\{
	\frac{s\Pi_z-\dot{\theta}_{5+m}}{\Omega_{0,s}}\left[\abs{\tilde\beta_0}^2-\abs{\tilde\alpha_0}^2\right]
	+ \frac{2m}{\Omega_{0,s}}\mathrm{Re}\left[\tilde\alpha_0^* \tilde\beta_0 e^{2i\Theta_{0,s}}\right]
	\right. \nonumber \\ &\left.
	+\sum_{n,\lambda'}
	\left[\sum_\lambda
	\left[
	\lambda\frac{s\Pi_z}{\Pi}\frac{\Pi+\lambda \dot{\theta}_{5+m}}{\Omega_\lambda}
	\left(\abs{\tilde\beta_{n,\lambda}^{(\lambda')}}^2 - \abs{\tilde\alpha_{n,\lambda}^{(\lambda')}}^2\right)
	+ \frac{2s\Pi_z}{\Pi}\frac{m}{\Omega_{n,\lambda}}
	\mathrm{Re}\left[\tilde{\alpha}_{n,\lambda}^{(\lambda')*}\tilde{\beta}_{n,\lambda}^{(\lambda')}\right]
	\right]
	\right. \right. \nonumber \\ &~~~~~~~~~~\left. \left.
	+\frac{m_B}{\Pi}\frac{m(\Omega_{n,+} - \Omega_{n,-} + 2\dot{\theta}_{5+m})}
	{\tilde{N}_n}
	\mathrm{Re}\left[e^{i(\Theta_{n,+} - \Theta_{n,-})}
	\left(\tilde{\beta}_{n,-}^{(\lambda')*} \tilde{\beta}_{n,+}^{(\lambda')}
	- \tilde{\alpha}_{n,+}^{(\lambda')*} \tilde{\alpha}_{n,-}^{(\lambda')}\right)\right]
	\right. \right. \nonumber \\ &~~~~~~~~~~\left. \left.
	+\frac{2m_B}{\Pi}\frac{(\Omega_{n,+} + \Pi + \dot{\theta}_{5+m})(\Omega_{n,-} + \Pi - \dot{\theta}_{5+m}) + m^2}
	{\tilde{N}_n}
	\mathrm{Re}\left[e^{i(\Theta_{n,+} + \Theta_{n,-})}\left(\tilde{\alpha}_{n,+}^{(\lambda')*}\tilde{\beta}_{n,-}^{(\lambda')}
	+ \tilde{\alpha}_{n,-}^{(\lambda')*}\tilde{\beta}_{n,+}^{(\lambda')}\right)\right]
	\right]
	\right\},
\end{align}
where
\begin{align}
	\tilde{N}_n = 
	\sqrt{\Omega_{n,+}\Omega_{n,-}(\Omega_{n,+} + \Pi + \dot{\theta}_{5+m})(\Omega_{n,-} + \Pi - \dot{\theta}_{5+m})}.
\end{align}
The axial charge is expressed as
\begin{align}
	q_5 &= \left.q_5\right\vert_{\mathrm{LLL}} +  \left.q_5\right\vert_{\mathrm{HLL}},
\end{align}
where
\begin{align}
	\left.q_5 \right\vert_{\mathrm{LLL}}
	&=
	\frac{g\abs{QB}}{4\pi^2}\int \dd p_z
	\left[
	\frac{s\Pi_z}{\Omega_0}\left(\abs{\beta_0}^2 - \abs{\alpha_0}^2\right)
	+ \frac{2m}{\Omega_0}\mathrm{Re}\left[\alpha_0^* \beta_0 e^{2i\Theta_0}\right] \right]
	\nonumber \\
	&=
	\frac{g\abs{QB}}{4\pi^2}\int \dd p_z
	\left[
	\frac{s\Pi_z-\dot{\theta}_{5+m}}{\Omega_{0,s}}
	\left(\abs{\tilde\beta_0}^2 - \abs{\tilde\alpha_0}^2\right)
	+ \frac{2m}{\Omega_{0,s}}\mathrm{Re}\left[\tilde\alpha_0^* \tilde\beta_0 e^{2i\Theta_{0,s}}\right] \right],
\end{align}
and
\begin{align}
	\left.q_5 \right\vert_{\mathrm{HLL}}
	&=
	\frac{g\abs{QB}}{4\pi^2}\int \dd p_z
	\sum_{n,\lambda}\left[
	\frac{m}{m_T}\frac{s\Pi_z}{\Omega_n}
	\left(
	\abs{\alpha_{n,1}^{(\lambda')}}^2 - \abs{\beta_{n,1}^{(\lambda')}}^2
	- \abs{\alpha_{n,2}^{(\lambda)}}^2 + \abs{\beta_{n,2}^{(\lambda)}}^2
	\right)
	\right. \nonumber \\ &~~~~~~~~~~~~~~~~~~~~~~~~~~~~~~~~~~~~~~~~\left.
	- \frac{2m_B}{m_T}\mathrm{Re}\left[\alpha_{n,1}^{(\lambda)*}\alpha_{n,2}^{(\lambda)}
	+\beta_{n,1}^{(\lambda)*}\beta_{n,2}^{(\lambda)}\right]
	-\frac{2m}{\Omega_n}\mathrm{Re}\left[\left(\alpha_{n,1}^{(\lambda)*}\beta_{n,1}^{(\lambda)}
	- \alpha_{n,2}^{(\lambda)*}\beta_{n,2}^{(\lambda)}\right)e^{2i\Theta_n}
	\right]
	\right]
	\nonumber \\
	&= 
	\frac{g\abs{QB}}{4\pi^2}\int \dd p_z
	\sum_{n,\lambda,\lambda'}
	\left[
	\frac{\lambda \Pi_B + \dot{\theta}_{5+m}}{\Omega_{n,\lambda}}
	\left(\abs{\tilde{\alpha}_{n,\lambda}^{(\lambda')}}^2 - \abs{\tilde{\beta}_{n,\lambda}^{(\lambda')}}^2\right)
	-\frac{2m}{\Omega_{n,\lambda}}
	\mathrm{Re}\left[
	\tilde{\alpha}_{n,\lambda}^{(\lambda')}\tilde{\beta}_{n,\lambda}^{(\lambda')}e^{2i\Theta_{n,\lambda}}
	\right]
	\right].
\end{align}
Finally the chiral mass operator is expressed as
\begin{align}
	\left\langle \bar{\psi} e^{2i\theta_m \gamma_5} i\gamma_5 \psi \right\rangle
	 &= \frac{g\abs{QB}}{2\pi^2}
	 \int \dd p_z\,\mathrm{Im}\left[
	 -\alpha_0^* \beta_0 e^{2i\Theta_0}
	 +\sum_{n,\lambda}
	 \left(\alpha_{n,1}^{(\lambda)*} \beta_{n,1}^{(\lambda)} 
	 - \alpha_{n,2}^{(\lambda)*}\beta_{n,2}^{(\lambda)}\right)e^{2i\Theta_{n}}
	 \right]
	 \nonumber \\
	 &=\frac{g\abs{QB}}{2\pi^2}
	 \int \dd p_z\,\mathrm{Im}\left[
	 -\tilde\alpha_0^* \tilde\beta_0 e^{2i\Theta_{0,s}}
	 +\sum_{n,\lambda,\lambda'}
	 \tilde{\alpha}_{n,\lambda}^{(\lambda')*}\tilde{\beta}_{n,\lambda}^{(\lambda')} e^{2i\Theta_{n,\lambda}}
	 \right].
\end{align}
Again we do not perform the regularization and renormalization of these quantities here.

%%%%%%%%%%%%%
\small
\bibliographystyle{utphys}
\bibliography{refs}
%%%%%%%%%%%%%
  
%%%%%%%%%%%%%
\end{document}